\documentclass[12pt]{article}
\usepackage{amsmath,amssymb,epsfig}
\makeatletter \@addtoreset{equation}{section} \makeatother
\renewcommand{\theequation}{\thesection.\arabic{equation}}
\newcommand{\ba}{\begin{array}}
\newcommand{\ea}{\end{array}}
\newcommand{\beq}{\begin{equation}}
\newcommand{\eeq}{\end{equation}}
\newcommand{\bea}{\begin{eqnarray}}
\newcommand{\eea}{\end{eqnarray}}

\def\bce{\begin{center}}
\def\ece{\end{center}}

\def\nonu{\nonumber}

\def\pa{\partial}
\def\al{\alpha}
\def\be{\beta}

\def\de{\delta}

\def\la{\lambda}

\def\eps6{{\displaystyle \mathop{\epsilon}^{6}}{}}

\def\nab6{{\displaystyle \mathop{\nabla}^{6}}{}}

\def\0{{\sst{(0)}}}
\def\1{{\sst{(1)}}}
\def\2{{\sst{(2)}}}
\def\3{{\sst{(3)}}}
\def\4{{\sst{(4)}}}
\def\5{{\sst{(5)}}}
\def\6{{\sst{(6)}}}
\def\7{{\sst{(7)}}}
\def\8{{\sst{(8)}}}

\def\ba{\begin{array}}
\def\ea{\end{array}}
\def\beq{\begin{equation}}
\def\eeq{\end{equation}}
\def\be{\begin{equation}}
\def\ee{\end{equation}}

\def\Tr{\mathop{\rm Tr}}

\def\la{\lambda}
\def\eps{\epsilon}

\def\ba{\begin{array}}
\def\ea{\end{array}}
\def\beq{\begin{equation}}
\def\eeq{\end{equation}}
\def\be{\begin{equation}}
\def\ee{\end{equation}}

\def\Tr{\mathop{\rm Tr}}

\def\la{\lambda}
\def\eps{\epsilon}

\newcommand{\bean}{\begin{eqnarray*}}
\newcommand{\eean}{\end{eqnarray*}}

\begin{document}
\thispagestyle{empty} \addtocounter{page}{-1}
   \begin{flushright}
\end{flushright}

\vspace*{1.3cm}
  
 \centerline{ \Large \bf  Are There Any New Vacua of }    
\vspace{.3cm} 
\centerline{ \Large \bf  
Gauged ${\cal N}=8$  Supergravity in Four Dimensions?     } 
\vspace*{1.5cm}
\centerline{\bf Changhyun Ahn {\rm and} Kyungsung Woo }
\vspace*{1.0cm} 
\centerline{\it  
Department of Physics, Kyungpook National University, Taegu
702-701, Korea} 
\vspace*{0.8cm} 
\centerline{\tt ahn@knu.ac.kr
} 
\vskip2cm

\centerline{\bf Abstract}
\vspace*{0.5cm}

We consider the most general $SU(3)$ singlet space of
gauged ${\cal N}=8$ supergravity in four-dimensions.
The $SU(3)$-invariant six scalar fields in the theory can be viewed in terms of 
six real four-forms. By exponentiating these four-forms, we eventually
obtain the new scalar potential. For the two extreme limits, 
we reproduce the previous results found by Warner in 1983. In
particular, for the ${\cal N}=1$ $G_2$ critical point, we find 
the constraint surface parametrized by three scalar fields on which 
the cosmological constant has the same value.
We obtain the BPS domain-wall solutions for restricted scalar
submanifold.
We also describe the three-dimensional mass-deformed 
superconformal Chern-Simons matter theory dual to 
the above supersymmetric
flows in four-dimensions.    

\baselineskip=18pt
\newpage
\renewcommand{\theequation}
{\arabic{section}\mbox{.}\arabic{equation}}

\section{Introduction}

The
three-dimensional ${\cal N}=6$ $U(N) \times U(N)$ 
Chern-Simons matter theory
with level $k$ 
has been constructed in \cite{ABJM}.
This theory can be regarded as the low energy limit of $N$ M2-branes at 
${\bf C}^4/{\bf Z}_k$ singularity.
The coupling of this theory may be thought of as $\frac{1}{k}$
and so this theory is weakly coupled for large $k$. 
On the other hand, for $k=1, 2$, the full ${\cal N}=8$ supersymmetry
is preserved and this theory becomes strongly coupled one. 

The renormalization group(RG) flow
between the ultraviolet fixed point and 
the infrared fixed point of the three-dimensional 
field theory can be obtained from gauged ${\cal N}=8$ 
supergravity theory in four-dimensions via AdS/CFT correspondence. 
The holographic
RG flow equations connecting ${\cal N}=8$ $SO(8)$ fixed point 
to ${\cal N}=2$ $SU(3) \times U(1)$ fixed point were found in 
\cite{AP,AW}(See also \cite{NW}) and
the other holographic
RG flow equations from ${\cal N}=8$ $SO(8)$ fixed point 
to
${\cal N}=1$ $G_2$
fixed point also were studied in 
\cite{AW,AI,AR99}(See also \cite{Warner83,AI02}).
The exact solutions to the $M$-theory lift of these
RG flows were constructed in \cite{CPW,AI}.

The mass deformed $U(2) \times U(2)$
Chern-Simons matter theory with level $k=1$ 
or $k=2$ preserving ${\cal N}=2$ $SU(3) \times U(1)$ symmetry was studied 
in \cite{Ahn0806n2,BKKS,KKM,KPR}  
and the corresponding mass deformation for ${\cal N}=1$ $G_2$
symmetric case was described in \cite{Ahn0806n1} and
moreover in \cite{Ahn0812} the nonsupersymmetric 
RG flow equations preserving $SO(7)^{\pm}$ have been discussed.  
Very recently, in \cite{BHPW}, the holographic
RG flow equations connecting ${\cal N}=1$ $G_2$ fixed point 
to ${\cal N}=2$ $SU(3) \times U(1)$ fixed point were found. 
Furthermore, the ${\cal N}=4$ and ${\cal N}=8$ supersymmetric flows have been 
studied in \cite{AW09}.

The gauged ${\cal N}=8$ supergravity in four-dimensions
has a scalar potential which
is a function of 70 scalars, in general \cite{CJ}. 
For one possible embedding of $SU(3)$ where 
the decompositions of three basic eight-representations of $SO(8)$
into $SU(3)$ representations are
$\bf 8_v, 8_s, 8_c \rightarrow 3 +\overline{3} + 1 + 1$, all of the 
35-dimensional representations of $SO(8)$ decompose into
$\bf 8 + 6 + \overline{6} + 3 +3 + \overline{3} +
\overline{3} + 1 + 1 + 1$. Then the set of 70 scalars in gauged ${\cal N}=8$
supergravity contains six singlets of $SU(3)$. 
For the other embeddings of $SU(3)$ where 
$\bf 8_v, 8_s, 8_c \rightarrow 8 $, 
all of the 
35-dimensional representations of $SO(8)$ decompose into
$\bf 27 + 8 $. This implies that there exist no $SU(3)$ singlets in the 
scalar sector.

It is known in \cite{Warner83} 
that $SU(3)$ singlet space with a breaking of the $SO(8)$ gauge group
into a group which contains $SU(3)$
may be written in terms of  the action of $SU(2) \times U(1)$ subgroup
of $SU(8)$ on 70-dimensional representation.
The $SU(2)$ group element is realized by usual three Euler angles.
Surprisingly, the scalar potential is independent of two of three
angles and depends on only four parameters. 

In this paper, we would like to revisit this $SU(3)$ sector of 
gauged ${\cal N}=8$ supergravity in four-dimensions.
Although there are six singlets of $SU(3)$, the known 
scalar potential is parametrized by only four.
There is a  room for the existence of two additional parameters leading to six
fields(which are the same number of $SU(3)$ singlets). 
Instead of possessing the
$SU(2)$ group as a subgroup, one can take any $2\times 2$ unitary 
matrix $U(2)$.  
Then this $U(2)$ group element is realized by four real parameters and
moreover, there are two real parameters(i.e., $U(1)$ angle and one real 
parameter): there exist six fields.

In section 2, we review the construction of scalar potential 
given in \cite{Warner83} for $SU(2) \times U(1)$ subgroup
of $SU(8)$ of gauged ${\cal N}=8$ supergravity.  
The main difference between this paper and the result of \cite{AW}
is that we take the scalar manifold with explicit two Euler angles.
Recall that in \cite{AW} we put those values to zero from the beginning.  
Although $A_1$ tensor of the theory depends on  one of the Euler
angles, after diagonalizing the $A_1$ tensor, this dependence
disappears and two eigenvalues play the role of superpotential for
a scalar potential which is the same as the one in \cite{Warner83}.

In section 3, 
we construct a new scalar potential for $U(2) \times U(1)$ subgroup
of $SU(8)$ of gauged ${\cal N}=8$ supergravity.
Although $A_1$ tensor of the theory depends on the parameters on
$U(2)$ group, 
after diagonalizing the $A_1$ tensor, the two eigenvalues 
become simple and they can be obtained  from the eigenvalues of section 2
by field redefinitions. Eventually one obtains a new scalar potential
which has also alternative form that can be read off from the scalar potential in
section 2.
We describe some properties of all the critical points behind 
this scalar potential and discuss some of the implications of our results.
We focus on the nontrivial supersymmetric critical points   
and obtain the BPS domain-wall solutions for restricted scalar
submanifold from direct extremization of energy-density.  
The three-dimensional mass-deformed 
Chern-Simons matter theory is also discussed. 

In section 4, we summarize what we obtain in this paper and 
make some comments on the future directions. 

In the Appendix, we present the detailed computations in section 2 and
section 3. 

\section{The $SU(3)$-invariant sector of  gauged ${\cal N}=8$ supergravity with
  four scalar space }

Let us define the complex coordinates $z_j$($j=1, 2, 3, 4$) 
on ${\bf R}^8$ and $x^j$($j=1, 2, \cdots, 8$)
is a set of cartesian coordinates in the vector representation of
$SO(8)$ of gauged ${\cal N}=8$ supergravity:
\bea
z_1 = x^1 + i x^2, \qquad 
z_2 = x^3 + i x^4, \qquad 
z_3 = x^5 + i x^6, \qquad
z_4 = x^7 + i x^8. 
\label{newbasis}
\eea
One takes the $SO(6)$ subgroup of $SO(8)$ which rotates $x^j$
($j=1, 2, \cdots, 6$) and stabilizes $x^7$ and $x^8$
and let $SU(3)$ be the subgroup of $SO(6)$ which rotates 
$z_j$($j=1, 2, 3$) and preserves the complex structure 
defined by
$\frac{i}{2} \sum_{j=1}^3  d z_j \wedge d
  \overline{z}_j $ on ${\bf R}^6$. The complex conjugation of $z_j$,
denoted by
$\overline{z}_j$, can be obtained as usual from (\ref{newbasis}).
One defines the two complex structures on the vector representation of 
$SO(8)$ as follows \cite{Warner83,BHPW}:
\bea
J^{\pm} =\frac{i}{2} \left( \sum_{j=1}^3  d z_j \wedge d
  \overline{z}_j \right)
\pm \frac{i}{2} d z_4 \wedge d \overline{z}_4.
\label{twoJ}
\eea

Then two of the scalar $SU(3)$ singlets can be obtained by 
$F_1^{\pm}$ with (\ref{twoJ}) 
while the remaining four of the scalar $SU(3)$ singlets are 
defined by $F_2^{\pm} + i F_3^{\pm}$ \cite{Warner83,BHPW}: 
\bea
F_1^{\pm}  & = & J^{\pm} \wedge J^{\pm}, \nonu \\
F_2^{+} + i F_3^{+}  & = & d z_1 \wedge d z_2 \wedge d z_3 \wedge d z_4,
\nonu \\
F_2^{-} + i F_3^{-} & = & 
d z_1 \wedge d z_2 \wedge d z_3 \wedge d \overline{z}_4.
\label{fourforms}
\eea
The $SU(3)$ subgroup leaves all these six forms invariant and there are
two $U(1)$'s which rotate $z_j$($j=1, 2, 3$) 
and $z_4$ respectively.
The $F_3^{\pm}$ are not invariant under these $U(1)$ actions.
These six real four-forms(half of them are self-dual and others are anti-self-dual) 
can be regarded as six scalar fields in gauged ${\cal N}=8$ supergravity
and they live in the coset \cite{BHPW}
\bea
\left[ \frac{SU(1,1)}{U(1)} \right]
\times \left[ \frac{SU(2,1)}{SU(2) \times U(1)} \right],
\label{manifolds}
\eea
as a submanifold of $\frac{E_{7(7)}}{SU(8)}$ which provides 70
real, physical scalars in gauged ${\cal N}=8$ supergravity.
Within $SO(8)$, the $SU(3)$ commutes with $U(1) \times U(1)$.
Within $SU(8)$, the $SU(3)$ commutes with $SU(2) \times U(1) \times U(1)$.
Within $E_{7(7)}$, the $SU(3)$ commutes with $SU(2,1) \times SU(1,1)$.
One parametrizes the scalar manifolds (\ref{manifolds}) 
using three complex scalar fields
$w_m$($m=1, 2, 3$)  with $E_{7(7)}$ components \cite{BHPW} as follows:
\bea
\phi_{ijkl} = \sum_{m=1}^3 \left[ (\mbox{Re} \; w_m) \; F_m^{+} +  (\mbox{Im}
\; w_m) \; F_m^{-} \right].
\label{para}
\eea

Let us consider the parametrization of \cite{Warner83,HW} for the $SU(3)$ singlet
space
\begin{eqnarray}
\phi_{ijkl} & = &  \; \la \; \cos\alpha\; F_1^{+} +
\la \; \sin \alpha\; F_1^{-} + \la' \; \cos \phi \;
\cos (\theta+ \psi) \; F_2^{+} 
 + \la' \; \sin \phi \; \cos(\theta-\psi) \; 
F_2^{-}  \; \nonu \\
& + & \la' \; \cos \phi \; \sin (\theta+\psi) \; F_3^{+}
-  \la' \; \sin \phi \; \sin (\theta-\psi) \; F_3^{-},
\label{para1}
\end{eqnarray}
where the six four-forms (\ref{fourforms}) are given explicitly by 
\begin{eqnarray}
  F_1^{\pm} &=& \varepsilon_{\pm} \left[ \; (\de^{1234}_{ijkl} \pm 
\de^{5678}_{ijkl})+
 (\de^{1256}_{ijkl} \pm \de^{3478}_{ijkl})+(\de^{3456}_{ijkl}
 \pm \de^{1278}_{ijkl}) \; \right],
\nonu \\
       F_2^{\pm} &=& \varepsilon_{\pm} \left[ -(\de^{1357}_{ijkl}
\pm \de^{2468}_{ijkl})+(\de^{2457}_{ijkl} \pm
\de^{1368}_{ijkl})+(\de^{2367}_{ijkl} \pm \de^{1458}_{ijkl}) +
 (\de^{1467}_{ijkl} \pm \de^{2358}_{ijkl}) \right], 
\nonu \\
 F_3^{\pm} &=& \varepsilon_{\pm} \left[ (\de^{2467}_{ijkl}
\mp \de^{1358}_{ijkl}) - (\de^{1367}_{ijkl} \mp
\de^{2458}_{ijkl}) - (\de^{1457}_{ijkl} \mp \de^{2368}_{ijkl}) -
 (\de^{2357}_{ijkl} \mp \de^{1468}_{ijkl}) \right],
\label{four}
\end{eqnarray} 
where $\varepsilon_{+}=1$ and $\varepsilon_{-}=i$ and $+$ gives
the scalars and $-$ gives the pseudoscalars \footnote{In \cite{HW}, the
similar construction of the scalar and pseudoscalar singlets of
$SU(3)$ is found. However, their $SU(3)$ rotates $z_j$( 
$j=2, 3, 4$) and therefore the nontrivial aspect appears in the first
two indices rather than the last two. This is the reason why the
explicit form for four-forms is different from each other. See also
the equation (3.1) of \cite{HW} corresponding to (\ref{four}). }.
Recall that the three angles $\phi, \theta$ and $\psi$
parametrize a general $SU(2)$ matrix in (\ref{manifolds}) and the
angle $\alpha$ parametrizes the $U(1)$ matrix in the first factor of
(\ref{manifolds}).
Originally, an arbitrary four-form, $\phi_{ijkl}$, is written as 
a product of $SU(8)$ element, which has four angle parameters, with seven canonical 
self-dual four forms that have   
two independent real parameters $\la$ and $\la'$.
This comes from the fact that  an arbitrary four-form, in principle, 
can be written as a product of $SU(8)$ element which has 63 parameters 
and 7 canonical self-dual four-forms leading to 
70 self-dual four-forms \cite{Warner84}.
It is easy to check (\ref{para1}) by performing the $SU(8)$ action 
on 70-dimensional representation in the space of self-dual complex four-forms.
One has the following relations \footnote{ 
Note that the general $SU(2)$ matrix element is given by
\bea
\left(
\begin{array}{cc}
e^{i\phi} \cos \theta \cos \psi - e^{-i\phi} \sin \theta \sin
\psi  & -e^{i\phi} \cos \theta \sin \psi - e^{-i\phi} \sin \theta \cos
\psi   \\
e^{i\phi} \sin \theta \cos \psi + e^{-i\phi} \cos \theta \sin
\psi &  -e^{i\phi} \sin \theta \sin \psi + e^{-i\phi} \cos \theta \cos
\psi   \\
\end{array}
\right) =
\left(
\begin{array}{cc}
\frac{w_2}{\la'} &  -\frac{w_3}{\la'}  \\
\frac{\overline{w}_3}{\la'} &   \frac{\overline{w}_2}{\la'}  \\
\end{array}
\right).
\label{su2}
\eea} 
between (\ref{para}) and (\ref{para1}):
\bea
w_1 & = & \la e^{ i \alpha}, 
\nonu \\
w_2 & = & 
\la' \left(e^{i\phi} \cos \theta \cos \psi - e^{-i\phi} \sin \theta \sin
\psi \right),
\nonu \\
w_3 & = & 
\la' \left(e^{i\phi} \cos \theta \sin \psi + e^{-i\phi} \sin \theta \cos
\psi \right).  
\label{relations}
\eea
Note that in \cite{AW}, we put the constraints $\theta=0=\psi$ at the
beginning, which lead to the vanishing of $F_3^{\pm}$ term in (\ref{para1}),
due to the fact that the scalar potential does not depend on
these values. In this paper, we relax these conditions and want to see how
those variables appear in the various places we describe. Actually these features
are crucial for describing the additional nonzero 
$F_3^{\pm}$ term in the next section 
when we turn on the more general four-forms. 

Let us contract the $\phi_{IJKL}$ (\ref{para1}) with the gamma matrices $\Gamma^I$.
We use $\Gamma^1={\bf 1}_{8\times 8}$ and $SO(7)$ gamma matrices
$\Gamma^J$
where $J=2, 3, \cdots, 8$ \cite{AW01}. 
Let us introduce gamma matrices
$\widetilde{\Gamma}^I=\left(\begin{array}{cc}
 0 &  {\bf 1}_{4 \times 4} \\
{\bf 1}_{4\times 4}  & 0\\
\end{array}\right) \Gamma^I \left(\begin{array}{cc}
 0 &  {\bf 1}_{4 \times 4} \\
{\bf 1}_{4\times 4}  & 0\\
\end{array}\right)$
and define the following quantity which was introduced in \cite{GW}
with new gamma matrices \bea
S_{AB} \equiv \phi_{IJKL} (\widetilde{\Gamma}^{IJKL})_{AB}.
\label{S}
\eea
Then one obtains the real part of (\ref{S})
as follows:
\bea
\left(
\begin{array}{ccc}
\la \cos \alpha {\bf 1}_{6 \times 6}  & {\bf 0}_{6\times 1} & {\bf
  0}_{6\times 1}  \\
{\bf 0}_{1\times 6} & -3\la \cos \alpha + 4 \la' \cos \phi \cos (\theta+\psi) 
 &   4 \la' \cos \phi \sin (\theta+\psi)   \\
{\bf 0}_{1 \times 6} &  4 \la' \cos \phi \sin (\theta+\psi) 
 & -3\la \cos \alpha -4 \la' \cos \phi \cos (\theta+\psi)   \\
\end{array} \right).
\label{realreal}
\eea
The supergravity scalar or pseudoscalar field corresponds to the mass terms of $(77)$
plus $(88)$ components as well as $(78)$ component for nonzero
$\theta$ and $\psi$ in the boundary
theory. 
The imaginary part of (\ref{S}) can be written as
\bea
\left(
\begin{array}{ccc}
 -3\la \sin \alpha - 4 \la' \sin \phi \cos (\theta-\psi) 
 &   4 \la' \sin \phi \sin (\theta-\psi)  & {\bf 0}_{1\times 6}
 \\
 -4 \la' \sin \phi \sin (\theta-\psi) 
 & -3\la \sin \alpha + 4 \la' \sin \phi \cos (\theta-\psi)  & 
{\bf 0}_{1\times 6}
  \\
{\bf 0}_{6 \times 1} & {\bf 0}_{6\times 1} & \la \sin \alpha {\bf 1}_{6 \times 6}
\end{array} \right).
\label{imaginary}
\eea
We want to see how the supergravity fields in (\ref{realreal}) and
(\ref{imaginary}) 
map onto
the corresponding boundary field theory objects. 

It is convenient to represent the six scalars with a $56 \times 56$
matrix,
the 56-bein, ${\cal V}(x)$:
\bea
{\cal V}(x)=
\mbox{exp} \frac{1}{2\sqrt{2}} \left(
\begin{array}{cc}
0 &  \phi_{ijkl}(x)  \\
 \phi^{ijkl}(x) & 0 
\end{array} \right).
\label{calV}
\eea  
The singlet $\phi_{ijkl}$ (\ref{para1}) is a $28 \times 28$ matrix and it consists of 
one $4 \times 4$ block diagonal matrix and three $8 \times 8$ block
diagonal matrices. For nonzero values for $\theta$ and $\psi$, the previous
six $4 \times 4$ block diagonal matrices \cite{AW} where
$\theta=0=\psi$ 
are generalized to
be the present three $8 \times 8$ block diagonal matrices.
Let us look at (\ref{calV}) closely.
One can select some matrix elements from $56 \times 56$ matrix of
(\ref{calV}) in the exponent. 
Let us choose the $(\mbox{row}, \mbox{column})$-elements
\bea
&& (1, 29), (1, 30), (1, 31), (1, 32), (2, 29), (2, 30), (2, 31), (2, 32), \nonu \\
&& (3, 29),
(3, 30), (3, 31), (3, 32), (4, 29), (4, 30), (4, 31), (4, 32),
\label{4x4}
\eea
which are the first $4 \times 4$ block diagonal of $\phi_{ijkl}$.
Similarly, one selects 
\bea
&& (29, 1), (29, 2), (29, 3), (29, 4), (30, 1), (30, 2), (30, 3), (30, 4), \nonu \\
&& 
(31, 1), (31, 2), (31, 3), (31, 4), (32, 1), (32, 2), (32, 3), (32, 4),
\label{4x4new}
\eea
which are the first $4 \times 4$ block diagonal of $\phi^{ijkl}$.
For the computation of (\ref{calV}), we need to know the multiple
product of the exponent. Then it is easy to see that 
when we compute the multiple product for the submatrix which consists
of (\ref{4x4}) and (\ref{4x4new}) with zero elements for other
components, the relevant matrix elements which contribute to the final
closed expression of (\ref{calV}) in the exponent are  given by
\bea
&& (1, 1), (1, 2), (1, 3), (1, 4), (2, 1), (2, 2), (2, 3), (2, 4), \nonu \\
&& (3, 1),
(3, 2), (3, 3), (3, 4), (4, 1), (4, 2), (4, 3), (4, 4),
\label{zero}
\eea
and the last one is
\bea
&& (29, 29), (29, 30), (29, 31), (29, 32), 
(30, 29), (30, 30), (30, 31), (30, 32), \nonu \\
&& 
(31, 29), (31, 30), (31, 31), (31, 32), 
(32, 29), (32, 30), (32, 31), (32, 32).
\label{anotherzero}
\eea
Therefore, one takes the $8\times 8$ block diagonal matrix 
as (\ref{zero}) for one-one block, (\ref{4x4}) for one-two block, 
(\ref{4x4new}) for two-one block and (\ref{anotherzero}) for two-two block.
After exponentiating this submatrix, then we redistribute them  
in $56 \times 56$ matrix and finally read off the 28-beins. 
In this way, one can also construct three $16 \times 16$ block
diagonal matrices. 

Then the exponent of (\ref{calV}) leads to a single $8 \times 8$ block diagonal
matrix and three  $16 \times 16$ block
diagonal matrices. 
Since four submatrices in the exponent are block diagonal in $56 \times 56$ matrix,
we do not need the nontrivial Baker-Hausdorff formulae here and the
above 56-bein (\ref{calV}) can be decomposed into four independent 
exponential factors where each exponent commutes with each other.   
Let us compute these factors explicitly.
First, one can easily get the $8 \times 8$ matrix by exponentiating 
a single $8 \times 8$ block diagonal
matrix in the first exponent after using the mathematica command ``MatrixExp''.
Secondly, in order to obtain the other factors of 56-bein, 
we compute the eigenvalues and eigenvectors for 
each $16 \times 16$ block
diagonal matrix in the last three exponents. 
Then one takes each $16\times 16$ matrix $U$ as a
transpose of eigenvectors. That is, each $U$ has 16 column vectors
that are $16 \times 1$ matrices respectively. 
Moreover, one obtains 
the $16 \times 16$ matrix $D$ easily by exponentiating 
a single $16 \times 16$ diagonal
matrix which is obtained from the eigenvalues using ``MatrixExp'' command.
Then the each $16 \times 16$ matrix component can be determined by 
$U D U^{-1}$ where $U^{-1}$ is an inverse matrix of $U$.
Also one can check that this $U D U^{-1}$ is equal to the original  
$16 \times 16$ block
diagonal matrix in the last three exponents.

Therefore, the 56-bein (\ref{calV}) can be written as
\bea
{\cal V}(x)=
\left(
\begin{array}{cc}
u_{ij}^{\;\;IJ}(x) & v_{ijKL}(x)  \\
v^{klIJ}(x) & u^{kl}_{\;\;KL}(x)
\end{array} 
\right),
\label{56bein}
\eea
where the 28-beins $ u^{IJ}_{\;\;KL}$ and $ v_{IJKL}$
are given in (\ref{uv}), (\ref{u}), (\ref{change}), (\ref{v}) and
(\ref{change1}) 
of the Appendix A.
Let us define $SU(8)$ T-tensor which is  
manifestly antisymmetric in the indices
$[ij]$ and $SU(8)$ covariant:
\bea
T_l^{\;kij} & = & 
\left(u^{ij}_{\;\;IJ} +v^{ijIJ} \right) \left( u_{lm}^{\;\;\;JK} 
u^{km}_{\;\;\;KI}-v_{lmJK} v^{kmKL} \right).
\label{ttens}
\eea
In particular, 
$A_1$ tensor is symmetric in $(ij)$
and $A_2$ tensor is antisymmetric in $[ijk]$ as follows: 
\bea
A_1^{\;\;ij} & =& 
-\frac{4}{21} T_{m}^{\;\;ijm}, \;\;\; A_{2l}^{\;\;\;ijk}=-\frac{4}{3}
T_{l}^{\;[ijk]}.
\label{a1a2}
\eea
The former appears in the variation of the gravitino of the theory
while the latter appears in the variation of 56 Majorana spinor of the theory.

It turns out that 
$A_1$ tensor has four distinct complex
values, $\hat{z}_1$, $\hat{z}_2$, $\hat{z}_3$ and $\hat{z}_4$ 
with degeneracies 6, 1, 1 and 2 respectively and has the following form
\bea
A_1^{\;IJ} = \left(
\begin{array}{cccccc|cc}
\hat{z}_1 &0 & 0& 0&0 &0 &0 & 0   \\
0 &\hat{z}_1 &0 &0 &0 &0 &0 & 0 \\
0 &0 &\hat{z}_1 &0 &0 &0 &0 &0   \\
0 &0 &0 &\hat{z}_1 &0 &0 &0 &0   \\
0 &0 &0 &0 &\hat{z}_1 &0 &0 &0   \\
0 &0 &0 & 0& 0&\hat{z}_1 &0 &0   \\
\hline
0 &0 &0 &0 &0 &0 &\hat{z}_2 &\hat{z}_4   \\
0 &0 &0 &0 &0 &0 &\hat{z}_4 &\hat{z}_3   \\
\end{array} \right), 
\label{A1A1}
\eea
where $\hat{z}_1$ is a function of $\la, \la', \alpha$ and $\phi$
and  $\hat{z}_2$, $\hat{z}_3$ and $\hat{z}_4$ depend on these as well
as $\theta$:
\bea
\hat{z}_{1} & = &
e^{-2i\left(\alpha+\phi\right)}\left[pq(e^{3i\alpha}p+q)r^{2}t^{2}+
e^{4i\phi}pq(e^{3i\alpha}p+q)r^{2}t^{2} \right. \nonu \\
 &    & + \left. e^{i\left(\alpha+2\phi\right)}(q^{3}+
2q(2+3q^{2})r^{2}t^{2}+e^{i\alpha}p(1+q^{2}+2(1+3q^{2})r^{2}t^{2}))
\right],
\nonu \\
\hat{z}_{2} & = & \frac{1}{4}e^{-2i\left(\alpha+2\phi\right)}
\left[4e^{2i\left(\alpha+2\phi\right)}(p^{3}+e^{3i\alpha}q^{3})
(1+t^{4})+4e^{2i\left(\theta+2\phi\right)}t^{2} \right. \nonu \\
 &  &  \times (2(p^{3}+e^{3i\alpha}q^{3})+6e^{i\alpha}
pq(p+e^{i\alpha}q)r^{2}\sin2\phi+(p^{3}+e^{3i\alpha}q^{3})t^{2}\cos4\phi\nonu
\\
 &    & -\left. i\cos2\theta(6e^{i\alpha}pq(p+e^{i\alpha}q)r^{2}
\sin2\phi+(p^{3}+e^{3i\alpha}q^{3})t^{2}\sin4\phi)) \right],
\nonu \\
\hat{z}_{3} & = & \frac{1}{4}e^{-2i\left(\alpha+2\phi\right)}
\left[4e^{2i\left(\alpha+2\phi\right)}(p^{3}+
e^{3i\alpha}q^{3})(1+t^{4})+4e^{2i\left(\theta+2\phi\right)}t^{2}\right.
\nonu
\\
 &  & \times(6e^{i\alpha}(p+e^{i\alpha}q)pqr^{2}(\cos2\phi+i\cos2\theta\sin2\phi)\nonu \\
 &  & + \left.
 (p^{3}+e^{3i\alpha}q^{3})(2+t^{2}(\cos4\phi+i\cos2\theta\sin4\phi)) \right],
\nonu \\
\hat{z}_{4} & = & 
\frac{1}{4}ie^{-2i\left(\alpha+2\phi\right)}(e^{4i\theta}-1)
(e^{4i\phi}-1)t^{2}\left[6e^{i\left(\alpha+2\phi\right)}pq(p+e^{i\alpha}q)r^{2}
\right. \nonu
\\
 &  & + \left. (p^{3}+e^{3i\alpha}q^{3})(1+e^{4i\phi})t^{2} \right].
\label{a1comp}
\eea
Here let us introduce 
the following quantities
\bea 
& & p \equiv \cosh \left(\frac{\la}{2\sqrt{2}}\right), \;\; q \equiv
\sinh\left(\frac{\la}{2\sqrt{2}}\right), 
\;\; r \equiv \cosh\left(\frac{\la'}{2\sqrt{2}}\right), \;\; t
\equiv \sinh\left(\frac{\la'}{2\sqrt{2}}\right). 
\label{pqrt}
\eea 
The $\hat{z}_1$ does not contain the field $\theta$ or $\psi$ and
coincides with the one by putting the constraints $\theta=0=\psi$ \cite{AW}.
It is easy to see that $\hat{z}_4$ becomes zero when $\theta=0$ due to
a factor $(e^{4i\theta}-1)$.
One obtains the two eigenvalues $z_2$ and $z_3$ for $2\times 2$ matrix with matrix
elements by $\hat{z}_2, \hat{z}_3$ and $\hat{z}_4$ in (\ref{A1A1}) and it turns out,
together with (\ref{a1comp}), 
that
\bea
z_2(\la,\la';\alpha,\phi) = \hat{z}_2|_{\theta=0}, \qquad \mbox{and} 
\qquad z_3(\la,\la';\alpha,\phi) = \hat{z}_3|_{\theta=0}.
\label{eigenvalues}
\eea
Furthermore, as noted in \cite{BHPW}, the eigenvalues $z_2$ and $z_3$
are related to each other by
\bea
z_3(\la,\la';\alpha,\phi) = z_2(\la,\la';\alpha,\phi)|_{\la'
  \rightarrow -\la', \phi \rightarrow -\phi}.
\label{z2z3}
\eea


Finally, 
the scalar potential from the general expression of \cite{dN82}
\bea
V= -g^2 \left( \frac{3}{4} \left| A_1^{\;ij} \right|^2-\frac{1}{24} \left|
A^{\;\;i}_{2\;\;jkl}\right|^2 \right), 
\nonu
\eea
can be written, by adding all the components of
$A_1, A_2$ tensors,  as
\bea 
&& V(\la, \la'; \al, \phi)  
= -g^2 \left[ \; \frac{3}{4} \times \left( 6|\hat{z}_1|^2 + |\hat{z}_2|^2 +
    |\hat{z}_3|^2 + 2|\hat{z}_4|^2
\right) \right. \nonu \\
&&  \left.  - \frac{1}{24} \times 6 \left( 4 \sum_{i=1}^4 |y_i|^2 
+12 \sum_{i=5 }^9 |y_i|^2 
+6|y_{10}|^2+
3 \sum_{i=11}^{12} |y_{i}|^2 
+ 6|y_{13}|^2\right) \; \right]
\nonu \\ 
&& =  \frac{1}{2} g^2 \left( s'^4 \left[ (x^2+3)c^3 + 4x^2v^3s^3 -
3v(x^2-1)s^3 + 12xv^2cs^2 - 6(x-1)cs^2 + 6(x+1)c^2sv \right] \right.  \nonu \\ 
&& \left. + 2s'^2 \left[ 
2(c^3+v^3s^3) + 3(x+1)vs^3 + 6xv^2cs^2 - 3(x-1)cs^2 -
6c \right]- 12c \right), 
\label{potential} 
\eea 
where we introduce
\bea & & c \equiv
\cosh\left(\frac{\la}{\sqrt{2}}\right), \; s \equiv \sinh\left(\frac{
\la}{\sqrt{2}}\right), 
\; c' \equiv
\cosh\left(\frac{\la'}{\sqrt{2}}\right), \; s' \equiv 
\sinh\left(\frac{\la'}{\sqrt{2}}\right), \nonu
\\ & & v \equiv \cos\alpha ,\;\;\;\;\;\;\;\;\;\;\; x \equiv 
\cos2\phi.   
\label{cs} 
\eea
Therefore, we explicitly show that the scalar potential does not
depend on the two angles $\theta$ and $\psi$ of $SU(2)$ matrix in 
(\ref{manifolds}).

As shown in \cite{AW}, one can write the scalar potential in terms of
a superpotential as follows: 
\bea 
V(\la, \la'; \al, \phi)= g^2 \left[ \frac{16}{3} \left|\frac{\partial
z_3}{\partial \la} \right|^2 + 4 \left|\frac{\partial z_3}{\partial
\la'} \right|^2  - 6  \left|z_3\right|^2 \right], 
\label{potpot} 
\eea
with a superpotential from (\ref{eigenvalues}), i.e., more explicitly, 
\bea 
z_3(\la, \la';\al, \phi) & = & 6e^{i(\alpha + 2\phi)}p^2qr^2t^2
 + 6e^{2i(\alpha + \phi)}pq^2r^2t^2  + p^3(r^4 + e^{4i\phi}t^4)  \nonu \\
& & +
e^{3i\alpha}q^3(r^4 + e^{4i\phi}t^4),
\label{z3superpotential}
\eea
where we use simplified notations of (\ref{pqrt}). 
Note that due to the property (\ref{z2z3}), when we replace $z_3$ with
$z_2$ in the right hand side of (\ref{potpot}), we obtain the same
expression for the scalar potential in terms of other superpotential 
$z_2$.  That is, there are two candidates for superpotential.

We expect to have the kinetic terms which contain 
$\theta$ and $\psi$ dependence as well as others.  
The resulting Lagrangian of scalar-gravity sector takes the form
\bea
\int d^4 x \sqrt{-g} \left[ \frac{1}{2} R  
- \frac{3}{8}\left(
\partial_{\mu} \lambda \right)^2 -\frac{3}{4} s^2 \left(
\partial_{\mu} \alpha \right)^2  -
\frac{1}{2}  \left( \partial_{\mu} \lambda' \right)^2 -
 s'^2 \left( \partial_{\mu} \phi \right)^2  -T- 
V(\la, \la'; \al, \phi) \right]
\label{action1}
\eea
with (\ref{cs}) and (\ref{potential})
where $\theta$ and $\psi$ dependent terms of kinetic energy are given by
\bea
T  & = &  
4r^{2}t^{2}(1+2t^{2})^{2} (\partial_{\mu}\psi)^2  +4  r^{2}t^{2}(1+
4r^2 t^2 \cos^2 2\phi) (\partial_{\mu}\theta)^2 
\nonu \\
& + &
8 r^{2} t^2(1+2t^{2})^{2}\cos 2\phi 
(\partial_{\mu}\theta) (\partial_{\mu}\psi).
\label{thetapsikin}
\eea
By substituting the usual domain-wall ansatz
into the Lagrangian (\ref{action1}), 
the Euler-Lagrangian equations for this  
are related to those for the energy functional $E$.
Here the energy-density per unit area transverse to $r$-direction is given
by reorganizing  the kinetic energy and the potential energy through 
usual squaring-procedure.
Since the remaining four kinetic terms plus the scalar potential in
(\ref{action1})
depend on only $\la, \la', \alpha$ and $\phi$ and do not depend on
$\theta$ and $\psi$, one can recombine those four kinetic terms
with the corresponding scalar potential parts in terms of the squares
of the four derivatives of superpotential as in \cite{AW}. Then we are
left with (\ref{thetapsikin}) which appears in the integrand of functional.    

How do we get the consistent BPS equations we have found in \cite{AW}
eventhough there exists the equation (\ref{thetapsikin})?
According to the observation of (\ref{kinother}), the above $T$ 
(\ref{thetapsikin})
can be written as $a_{+} G_{+}^2 + a_{-} G_{-}^2$. 
In order to have the
BPS bound, inequality of the energy-density, one should have $G_{\pm}
=0$
where $a_{\pm}$ and $G_{\pm}$ are defined through (\ref{kinother}). 
Then it is easy to
check that the solutions for these provide $\pa_{r} \theta =0 =\pa_{r}
\psi$.  Therefore, one can take $\theta=0=\psi$ which are the same 
as the expectation
values of ${\cal N}=8$ $SO(8)$ maximal supersymmetric case, along the
whole RG  flow. 

For the supersymmetry counting, one considers, for example,  
the variation of spin 
$\frac{3}{2}$ field. 
The variation of this gravitino contains gravitino, supersymmetry
parameter and the component of $A_1$ tensor. Recall that 
one should multiply the $2\times 2$ matrix $M$, 
obtained from the eigenvectors in the submatrix of nondiagonal 
$A_1$ tensor(characterized by $\hat{z}_2, \hat{z}_3, \hat{z}_4$ in (\ref{a1comp})),
and its inverse $M^{-1}$ like as $\widetilde{A}_1=M A_1 M^{-1}$ in 
(\ref{eigenvalues}), 
to diagonalize it.  
Moreover, one gets the corresponding transformed gravitino $\widetilde{\psi}^i$ and
supersymmetry parameter $\widetilde{\epsilon}^i$. 
After this procedure, one can go through the steps we
did in \cite{AW} and it is clear that there exists the supersymmetric
bosonic background consistent with the above BPS bound.

\section{
The $SU(3)$-invariant sector of  gauged ${\cal N}=8$ supergravity
 with six scalar space}

Let us introduce three complex fields
\bea
w_1 =\la e^{i\alpha}, \qquad
w_2 =\la' e^{i \phi}, \qquad
w_3 = \rho e^{i \varphi}
\label{w1w2w3}
\eea
instead of (\ref{relations}).
In terms of $2\times 2$ $U(2)$ group element,
$w_2$ and $w_3$ are located at 
\bea
\left(
\begin{array}{cc}
\la' e^{i\phi}  & -\rho e^{i\varphi}   \\
\rho e^{-i\varphi} & \la' e^{-i\phi}   \\
\end{array}
\right)
\nonu
\eea
compared with the one in (\ref{su2}) in previous section.
Let us consider more general $SU(3)$ singlet space parametrized by
six fields
\begin{eqnarray}
\phi_{ijkl}  =   \la \cos \alpha F_1^{+} +
\la  \sin \alpha F_1^{-} + \la'  \cos \phi  F_2^{+} 
 + \la' \sin \phi  F_2^{-}   + \rho \cos \varphi   F_3^{+}
+  \rho \sin \varphi   F_3^{-},
\label{phiijkl1}
\end{eqnarray}
where six four-forms are given in (\ref{four}).
Starting from (\ref{calV}), one can read off  
a single $8 \times 8$ block diagonal
matrix and three  $16 \times 16$ block
diagonal matrices from (\ref{phiijkl1}). 
Then the corresponding $8 \times 8$ matrix and three 
$16\times 16$ matrices can be obtained by using the exponentiating 
procedure we described in previous section exactly.
Then one arrives at the 56-bein given by (\ref{56bein}).
The complete expression for $u$ and $v$
is given in (\ref{uv1}), (\ref{u1}), (\ref{Change}), (\ref{v1}) and
(\ref{Change1}) 
of Appendix B.

The real part of (\ref{S}) can be computed as follows:
\bea
\left(
\begin{array}{ccc}
\la \cos \alpha {\bf 1}_{6 \times 6}  & {\bf 0}_{6\times 1} & {\bf
  0}_{6\times 1}  \\
{\bf 0}_{1\times 6} & -3\la \cos \alpha + 4 \la' \cos \phi 
 &   4 \rho \cos \varphi   \\
{\bf 0}_{1 \times 6} &  4 \rho \cos \varphi  
 & -3\la \cos \alpha -4 \la' \cos \phi    
\end{array} \right),
\label{S1}
\eea
which is traceless. Note that there exist the off-diagonal terms due
to the factor $\rho \cos \varphi$ which vanishes as $\rho$ goes to zero. 
The imaginary part of (\ref{S}) is given by
\bea
\left(
\begin{array}{ccc}
 -3\la \sin \alpha - 4 \la' \sin \phi  &   
-4 \rho \sin \varphi  & {\bf 0}_{1\times 6}
 \\
 -4 \rho \sin \varphi 
 & -3\la \sin \alpha +4 \la' \sin \phi   & 
{\bf 0}_{1\times 6}
 \\
{\bf 0}_{6 \times 1} & {\bf 0}_{6\times 1} & \la \sin \alpha {\bf 1}_{6 \times 6}
\end{array} \right)
\label{imagS1}
\eea
which is also traceless and the off-diagonal terms arise 
in $(12)$-component and $(21)$-component and they vanish as $\rho$
goes to zero.
We would like to see how the supergravity fields (\ref{S1}) and (\ref{imagS1}) map onto
the corresponding boundary field theory objects in this
parametrization also and the correspondence will appear at the end of
this section. 

The
$A_1$ tensor has four distinct complex
values, $\hat{z}_1$, $\hat{z}_2$, $\hat{z}_3$ and $\hat{z}_4$ 
with degeneracies 6, 1, 1 and 2 respectively and has the following form
\bea
A_1^{\;IJ} = \left(
\begin{array}{cccccc|cc}
\hat{z}_1 &0 & 0& 0&0 &0 &0 & 0   \\
0 &\hat{z}_1 &0 &0 &0 &0 &0 & 0  \\
0 &0 &\hat{z}_1 &0 &0 &0 &0 &0   \\
0 &0 &0 &\hat{z}_1 &0 &0 &0 &0   \\
0 &0 &0 &0 &\hat{z}_1 &0 &0 &0   \\
0 &0 &0 & 0& 0& \hat{z}_1 &0 &0   \\
\hline
0 &0 &0 &0 &0 &0 &\hat{z}_2 &\hat{z}_4   \\
0 &0 &0 &0 &0 &0 &\hat{z}_4 &\hat{z}_3   \\
\end{array} \right), 
\label{A1}
\eea
where $\hat{z}_1,  \hat{z}_2, \hat{z}_3$ and $\hat{z}_4$  
are functions of $\la, \la', \rho, \alpha, \phi$ and $\varphi$ as follows.
The diagonal terms in (\ref{A1}) are $(11), (22), (33), (44), (55)$
and $(66)$-component
\bea
\hat{z}_{1} & = & \frac{e^{-2i\alpha}}{\rho^{2}+\la'^{2}}\left[e^{i\alpha}
(e^{i\alpha}p(1+2r^{2}t^{2}+q^{2}(1+6r^{2}t^{2})) 
  + 
q(4r^{2}t^{2}+q^{2}(1+6r^{2}t^{2})))
(\rho^{2}+\la'^{2}) \right. \nonu \\
 & + & \left. 2pq(q+e^{3i\alpha}p)r^{2}t^{2}(\la'^{2}
\cos 2\phi+\rho^{2}\cos 2\varphi )\right],
\label{z1hat}
\eea
and the $(77)$-component 
\bea
\hat{z}_{2} & = & \frac{e^{-4i(\phi+\varphi)}}{(\rho^{2}+\la'^{2}
)^{2}}\left[e^{2i\phi}(e^{i(3\alpha+2\phi)}q^{3}t^{4}\rho^{4}+
e^{2i\phi}p^{3}t^{4}\rho^{4}+e^{3i(\alpha+2\varphi)}q^{3}t^{4}\rho^{2}
\la'^{2} + 
 e^{i(3\alpha+2\varphi)}q^{3}t^{4}\rho^{2}\la'^{2} \right. \nonu \\
& + & e^{2i\varphi}p^{3}t^{4}
\rho^{2}\la'^{2}+e^{6i\varphi}p^{3}t^{4}\rho^{2}\la'^{2} + 
 6e^{2i(\alpha+\phi+\varphi)}pq^{2}r^{2}t^{2}\rho^{2}(\rho^{2}  + 
\la'^{2}) \nonu \\
& + & 6e^{i(\alpha+2(\phi+\varphi))}p^{2}qr^{2}t^{2}\rho^{2}
(\rho^{2}+\la'^{2})+  
 6e^{2i(\alpha+2\varphi)}pq^{2}r^{2}t^{2}\la'^{2}(\rho^{2}+\la'^{2}
) \nonu \\
 & + & 6e^{i(\alpha+4\varphi)}p^{2}qr^{2}t^{2}
\la'^{2}(\rho^{2}+\la'^{2}) + 
 e^{i(3\alpha+2\phi+4\varphi)}q^{3}(\rho^{4}+t^{4}\rho^{4}+2\rho^{2}
\la'^{2}+\la'^{4}+2t^{2}(\rho^{2}+\la'^{2})^{2})\nonu
\\
 & + &
 e^{2i(\phi+2\varphi)}p^{3}(\rho^{4}+t^{4}\rho^{4}+2\rho^{2}
\la'^{2}+\la'^{4}+2t^{2}(\rho^{2}+\la'^{2})^{2}))\nonu
\\
 & + &
 \left. 2e^{2i(\phi+\varphi)}(p^{3}+e^{3i\alpha}q^{3})t^{4}
\la'^{2}(e^{2i\phi}\rho^{2}+e^{2i\varphi}\la'^{2})\cos 2\phi
\right],
\label{2}
\eea
and $(88)$-component
\bea
\hat{z}_{3} & = & \frac{e^{-2i\varphi}}{(\rho^{2}+\la'^{2})^{2}}
\left[e^{2i\varphi}p\rho^{4}+e^{2i\varphi}pq^{2}\rho^{4}+e^{i(3\alpha+
2\varphi)}q^{3}\rho^{4} + 
 2e^{2i\varphi}p\rho^{2}\la'^{2}+2e^{2i\varphi}pq^{2}\rho^{2}\la'^{2}
\right. \nonu \\ 
& + &
2e^{i(3\alpha+2\varphi)}q^{3}\rho^{2}\la'^{2} + 
 e^{2i\varphi}p\la'^{4}+e^{2i\varphi}pq^{2}\la'^{4}+e^{i(3\alpha+2\varphi)}q^{3}
\la'^{4}+2e^{2i\varphi}t^{2}(\rho^{2}+\la'^{2})(p(\rho^{2}+
\la'^{2})\nonu \\
 & + & pq^{2}(\rho^{2}+3e^{2i(\alpha+\varphi)}\rho^{2}+\la'^{2}+
3e^{2i(\alpha+\phi)}\la'^{2})+e^{i\alpha}q(e^{2i\alpha}q^{2}
(\rho^{2}+\la'^{2})\nonu \\
 & + &  3p^{2}(e^{2i\varphi}\rho^{2}+e^{2i\phi}
\la'^{2})))+t^{4}(e^{2i\varphi}\rho^{2}+e^{2i\phi}
\la'^{2})((p^{3}+e^{3i\alpha}q^{3})\rho^{2} 
 +  e^{4i\varphi}(p^{3}+e^{3i\alpha}q^{3})\rho^{2} \nonu \\
&+& \left.
6e^{i(\alpha+2\varphi)}pq(p+e^{i\alpha}q)(\rho^{2}+\la'^{2}) 
+  2e^{2i\varphi}(p^{3}+e^{3i\alpha}q^{3})
\la'^{2}\cos 2\phi)\right].
\label{3}
\eea 
There exist nondiagonal terms, $(78)$-component and $(87)$-component,  
\bea
\hat{z}_{4} & = &
\frac{e^{-i(\phi+3\varphi)}}{(\rho^{2}+\la'^{2})^{2}}
(-e^{2i\phi}+e^{2i\varphi})t^{2}\rho\la'\left[(p^{3}+
e^{3i\alpha}q^{3})t^{2}\rho^{2}
 +  e^{4i\varphi}(p^{3}+e^{3i\alpha}q^{3})t^{2}\rho^{2} \right.
\nonu \\
&+ & \left.
6e^{i(\alpha+2\varphi)}pq(p+e^{i\alpha}q)r^{2}(\rho^{2}+
\la'^{2})
  +  2e^{2i\varphi}(p^{3}+e^{3i\alpha}q^{3})t^{2}
\la'^{2}\cos 2\phi \right],
\label{4}
\eea
which vanishes for $\rho=0$ or $\la'=0$ which corresponds to the fact
that $\hat{z}_4$ vanishes when $\theta=0$ in section 2. 
Here we introduce the hyperbolic functions, for simplicity, as in
previous section
\bea 
p \equiv \cosh \left(\frac{\la}{2\sqrt{2}}\right), q \equiv
\sinh\left(\frac{\la}{2\sqrt{2}}\right), 
r \equiv \cosh\left(\frac{\sqrt{\la'^2+\rho^2}}{2\sqrt{2}}\right),
t
\equiv \sinh\left(\frac{\sqrt{\la'^2+\rho^2}}{2\sqrt{2}}\right). 
\label{pqrt1}
\eea 
Compared to (\ref{pqrt}), the first two are the same and the last two
are different in the sense that $\la'$ is replaced by $\sqrt{\la'^2 +\rho^2}$. 

One obtains the two eigenvalues $z_2$ and $z_3$ for $2\times 2$ matrix with matrix
elements by $\hat{z}_2, \hat{z}_3$ and $\hat{z}_4$ in (\ref{2}),
(\ref{3}), and 
(\ref{4}) and it turns out
that \footnote{We emphasize here the notations 
for $z_3(\la, \la', \rho; \al, \phi,
  \varphi)$ and 
$z_3(\la, \la'; \al, \phi)$. The former is parametrized by six
fields (\ref{superpotential2}) 
and the latter is parametrized by four fields
(\ref{z3superpotential}). The other forms for these can be written as
$z_2 =\frac{1}{2} \left( \hat{z}_2 +\hat{z}_3 -
\sqrt{\hat{z}_2^2+4\hat{z}_4^2-2\hat{z}_2
  \hat{z}_3+\hat{z}_3^2}\right)$ and $z_3 =\frac{1}{2} \left(
\hat{z}_2 +
\hat{z}_3 +
\sqrt{\hat{z}_2^2+4\hat{z}_4^2-2\hat{z}_2
  \hat{z}_3+\hat{z}_3^2}\right)$ in the eigenvalue calculation. One
can check that these are equal to (\ref{superpotential2}). }
\bea
z_2(\la, \la', \rho; \al, \phi, \varphi)   
& = & z_2(\la, \la'; \al, \phi)|_{\la' \rightarrow \sqrt{\la'^2 +\rho^2},\; 
  \phi \rightarrow \frac{1}{2} \cos^{-1} \left[
\frac{\la'^2 \cos2\phi + \rho^2 \cos2\varphi}{\la'^2+\rho^2} \right]}, 
\nonu \\
z_3(\la, \la', \rho; \al, \phi, \varphi)   
& = & z_3(\la, \la'; \al, \phi)|_{\la' \rightarrow \sqrt{\la'^2 +\rho^2},\; 
  \phi \rightarrow \frac{1}{2} \cos^{-1} \left[
\frac{\la'^2 \cos2\phi + \rho^2 \cos2\varphi}{\la'^2+\rho^2} \right]},
\label{superpotential2}
\eea
where $z_2(\la, \la'; \al, \phi)$ and 
$z_3(\la, \la'; \al, \phi)$ are 
the two eigenvalues of previous section (\ref{eigenvalues}) and
two candidates for the superpotential.
Furthermore, 
the eigenvalues $z_2$ and $z_3$
for six fields are related to each other by
\bea
z_3 = z_2|_{\sqrt{\la'^2 +\rho^2}
  \rightarrow -\sqrt{\la'^2 +\rho^2},  \frac{1}{2} \cos^{-1} \left[
\frac{\la'^2 \cos2\phi + \rho^2 \cos2\varphi}{\la'^2+\rho^2} \right]
\rightarrow -\frac{1}{2} \cos^{-1} \left[
\frac{\la'^2 \cos2\phi + \rho^2 \cos2\varphi}{\la'^2+\rho^2} \right]}.
\label{Relation}
\eea
This feature occurs also in (\ref{z2z3}) of previous section.
Of course, the eigenvalue $\hat{z}_1$ (\ref{z1hat}) can be obtained from 
the $\hat{z}_1$ (\ref{a1comp}) in section 2 by same transformation as 
(\ref{superpotential2}).
From the relations (\ref{relations}), by replacing
$\la'$ and $\phi$ of section 2 with $\chi$ and $\beta$
respectively, 
one has
\bea
\chi^2  & = & |w_2|^2 +|w_3|^2 =\la'^2 + \rho^2, \nonu \\
\cos 2 \beta & = & \frac{|w_2|^2 \cos (2\mbox{Arg}\;w_2) + |w_3|^2 \cos
  (2\mbox{Arg}\;w_3)}{|w_2|^2+|w_3|^2} =
 \frac{\la'^2 \cos 2\phi + \rho^2 \cos 2\varphi}{\la'^2+\rho^2},
\label{concon}
\eea
where we use (\ref{w1w2w3}).

The scalar potential can be obtained from the general procedure, like
as in (\ref{potential}), and it leads to
\bea
&& V(\la, \la', \rho; \al, \phi, \varphi)  
 =  
\nonu \\
&& \frac{1}{(\rho^{2}+\la'^{2})^{2}}2
\left[(1+2q^{2})((-3+8(-1+5p^{2}q^{2}
)r^{2}t^{2} 
 +2(7+76p^{2}q^{2})r^{4}t^{4})
\rho^{4} \right. \nonu \\
&& +  2(-3+4r^{2}t^{2}(-2+3r^{2}t^{2} 
 +2p^{2}q^{2}(5+18r^{2}t^{2})))
\rho^{2}\la'^{2}+(-3+8(-1+5p^{2}q^{2})r^{2}t^{2}
\nonu \\
 &  & +2(7+76p^{2}q^{2})r^{4}t^{4})
\la'^{4})+2r^{2}t^{2}(12pq(\rho^{2}+\la'^{2})
\cos\alpha ((2r^{2}t^{2} 
 +3q^{4}(1+2t^{2})^{2} \nonu \\
&& +3(q+2qt^{2})^{2}
)(\rho^{2}+\la'^{2})+2(r^{2}t^{2}+q^{4}
(1+2t^{2})^{2} 
 +(q+2qt^{2})^{2})(\la'^{2}
\cos2\phi+\rho^{2}\cos2\varphi))\nonu \\
 &  &
 +4p^{3}q^{3}\cos3\alpha ((\rho^{2}+\la'^{2})^{2}+
4r^{2}t^{2}(\la'^{2}\cos2\phi+\rho^{2}
\cos2\varphi)^{2})\nonu \\
 &  & +(1+2q^{2})(12p^{2}(q+2qt^{2})^{2}
(\rho^{2}+\la'^{2})\cos2\alpha(\la'^{2}
\cos2\phi+\rho^{2}\cos2\varphi) \nonu \\
 &  & \left. +r^{2}(t+2q^{2}t)^{2}
(\la'^{4}\cos4\phi+4\rho^{2}\la'^{2}\cos2\phi
\cos2\varphi+\rho^{4}\cos4\varphi)))\right],
\label{Poten}
\eea
with (\ref{pqrt1}).
Although the structure of this scalar potential seems to be very complicated, 
it is obvious that this leads to the following result, by simply taking
the subtraction between these two, 
\bea 
&& V(\la, \la', \rho; \al, \phi, \varphi)= \nonu \\ 
&&   \frac{1}{2} g^2 \left( s'^4 \left[ (x^2+3)c^3 + 4x^2v^3s^3 -
3v(x^2-1)s^3 + 12xv^2cs^2 - 6(x-1)cs^2 + 6(x+1)c^2sv \right] \right.  \nonu \\ 
&& \left. + 2s'^2 \left[ 
2(c^3+v^3s^3) + 3(x+1)vs^3 + 6xv^2cs^2 - 3(x-1)cs^2 -
6c \right]- 12c \right), 
\label{potential1} 
\eea 
where the redefined quantities are given by 
\bea & & c \equiv
\cosh\left(\frac{\la}{\sqrt{2}}\right), \; s \equiv \sinh\left(\frac{
\la}{\sqrt{2}}\right), 
\; c' \equiv
\cosh\left(\frac{\sqrt{\la'^2+\rho^2}}{\sqrt{2}}\right), \; s' \equiv 
\sinh\left(\frac{\sqrt{\la'^2 + \rho^2}}{\sqrt{2}}\right), \nonu
\\ & & v \equiv \cos\alpha ,\;\;\;\;\;\;\;\;\;\;\; x \equiv \cos
2\beta \equiv
\frac{\la'^2 \cos2\phi + \rho^2 \cos2\varphi}{\la'^2+\rho^2}.   
\label{csrelation} 
\eea
Compared to (\ref{cs}), the first two are the same, the next two
are different in the sense that $\la'$ is replaced by 
$\sqrt{\la'^2 +\rho^2}$ and the last one has nontrivial expression.
If $\rho$ goes to zero, then $x$ in (\ref{csrelation}) becomes
$x$ in (\ref{cs}) or if $\varphi=\phi$, then  one has a coincidence
between two expressions.
One sees the symmetry between $(\la',\phi)$ and $(\rho,\varphi)$
behind this scalar potential. That is, the scalar potential is
invariant under $\la' \leftrightarrow \rho$ and 
$\phi \leftrightarrow \varphi$.
In other words, the critical points for $\rho=0$ are
exactly the same as those for $\la'=0$.
From the explicit form for the four-form fields (\ref{four}),
there is a symmetry between the index 7 and the index 8.
If we exchange them, then $F_{2}^{\pm}$ goes to $F_{3}^{\pm}$ up to
signs and vice versa. 
By comparing (\ref{potential1}) with (\ref{potential}), one realizes that
there exists simple relation between the potential (\ref{potential1}) here and the
potential (\ref{potential}) appeared in previous section. More explicitly,
one obtains \footnote{Let us denote $V(\la, \la', \rho; \al, \phi,
  \varphi)$ by either (\ref{Poten}) or (\ref{potential1}) parametrized
  by six fields and 
$V(\la, \la'; \al, \phi)$ by (\ref{potential}) parametrized by four
fields as we write the arguments explicitly. }
\bea 
V(\la, \la', \rho; \al, \phi, \varphi)  =  
V(\la, \la'; \al, \phi)|_{\la' \rightarrow \sqrt{\la'^2 +\rho^2},\; 
  \phi \rightarrow \frac{1}{2} \cos^{-1} \left[
\frac{\la'^2 \cos2\phi + \rho^2 \cos2\varphi}{\la'^2+\rho^2} \right]}, 
\label{Rel2}
\eea
where the field transformations in the right hand side is exactly the
same as the one in (\ref{concon}). 
Although the components of $A_1$ and $A_2$ tensors are not transformed
from those in section 2 via (\ref{concon}), the scalar potential,
coming from the sum of the squares of absolute value of these,
is transformed via (\ref{concon}) and is summarized by (\ref{Rel2}).

From the first derivatives of scalar potential 
$V(\la, \la', \rho; \al, \phi, \varphi)$ 
with six fields for the critical points
\bea
\frac{\pa V}{\pa \la}  =  0, \qquad \frac{\pa V}{\pa \la'} =0, \qquad 
\frac{\pa V}{\pa \rho} =0, \qquad
\frac{\pa V}{\pa \alpha}  =  0, \qquad
\frac{\pa V}{\pa \phi} =0, \qquad
\frac{\pa V}{\pa \varphi} =0, 
\label{firstder}
\eea
it is easy to show that there exists, for the scalar field $V(\la, 
\la', \rho; \al, \phi, \varphi)$,  
\bea
\frac{\la'}{\sqrt{\la'^2+\rho^2}} \left(\frac{\pa V}{\pa \la'}\right) +
\frac{\rho}{\sqrt{\la'^2+\rho^2}} \left(\frac{\pa V}{\pa \rho}\right) =0.
\label{rel}
\eea 
The left hand side in (\ref{rel}) coincides with the expression of
$\frac{\pa V(\la, \la'; \al, \phi)}{\pa \la'}$ with the substitutions 
$\la' \rightarrow \sqrt{\la'^2 +\rho^2}$ and $ 
  \phi \rightarrow \frac{1}{2} \cos^{-1} \left[
\frac{\la'^2 \cos2\phi + \rho^2 \cos2\varphi}{\la'^2+\rho^2} \right]$
as in (\ref{concon}).  
Similarly, one obtains a relation,  for the scalar field $V(\la, 
\la', \rho; \al, \phi, \varphi)$,
\bea
-\frac{\la'^2+\rho^2}{2\la'^2 \sin 2\phi} \left(\frac{\pa V}{\pa \phi}\right) 
+\frac{\la'^2+\rho^2}{2\rho^2 \sin 2\varphi} \left(\frac{\pa V}{\pa \varphi}\right)=0.
\label{rel1}
\eea
In this case, the left hand side in (\ref{rel1}) is equal to the expression of
$\frac{\pa V(\la, \la'; \al, \phi)}{\pa \cos 2\phi}$ with the above
replacements for $\la'$ and $\phi$ as in (\ref{rel}). 
Of course, it is obvious that (\ref{rel}) and (\ref{rel1}) are
satisified automatically if (\ref{firstder}) is satisfied.
It is obvious that 
the first relation of (\ref{firstder}) is the same as 
the expression of
$\frac{\pa V(\la, \la'; \al, \phi)}{\pa \la}$ while 
the fourth relation of (\ref{firstder}) is equal to 
the expression of
$\frac{\pa V(\la, \la'; \al, \phi)}{\pa \alpha}$, with the above
replacements for $\la'$ and $\phi$ after doing these derivatives.
Therefore, 
from the critical point conditions for four scalar fields, 
\bea
\pa_{\la} V(\la, \la'; \al, \phi)=
\pa_{\alpha} V(\la, \la'; \al, \phi)=
\pa_{\la'} V(\la, \la'; \al, \phi)=
\pa_{\phi} V(\la, \la'; \al, \phi)=0,
\label{fourf}
\eea
one obtains the corresponding critical point conditions for six
fields.
In other words, instead of using (\ref{firstder}) directly, 
one uses some linear combinations of them (\ref{rel}) and (\ref{rel1}).
Then we need to check whether there are further constraints on the six
fields as well as the results from (\ref{fourf}) 
by looking at (\ref{firstder}) and exhausting the remaining
vanishing conditions.

Now we describe the critical points for the scalar potential 
(\ref{Poten}) we have
found by looking at those for the scalar potential (\ref{potential})  
and finding the further constraints on the remaining fields.  

$\bullet$ Nonsupersymmetric $SO(7)^{-}$ critical point 

One \cite{Englert} has the following critical values
\bea
\la=\sqrt{2}\sinh^{-1}\left(\frac{1}{2}\right)=\sqrt{\la'^2+\rho^2}, 
\qquad \alpha=\frac{\pi}{2}=\frac{1}{2} \cos^{-1} \left(
\frac{\la'^2 \cos2\phi + \rho^2 \cos2\varphi}{\la'^2+\rho^2} \right),
\label{cond}
\eea
after solving the four conditions characterized by    
(\ref{rel}), (\ref{rel1}) and the first and the fourth of (\ref{firstder}).  
We explicitly write the $\beta$ given in (\ref{csrelation}) here in order to
emphasize the existence of the number of independent fields above. 
Now we have to check other conditions in the sense that we are 
dealing with six fields.
Then it is straightforward to check the remaining second, third, fifth
and sixth equations of
(\ref{firstder}) satisfy under the condition
$
\la' \cos \phi =0$.
This implies, under the condition for nonzero $\la'$(When $\la'=0$,
then we end up with the case which was discussed in section 2), that 
\bea
\phi =\frac{\pi}{2} =\varphi,
\label{extra}
\eea
by eliminating the $\rho$ field from two equations (\ref{cond}).
Therefore, the additional field $\varphi$ is fixed as the $\phi$ field 
while the $\la'$ field is not fixed but the square sum $\la'^2 +\rho^2$
is fixed according to (\ref{cond}).
This result is summarized in Table 1 where the five fields among six
fields are fixed.
Since the angles $\alpha, \phi$ and $\varphi$ are equal to
$\frac{\pi}{2}$ via (\ref{cond})  and (\ref{extra}) and from (\ref{phiijkl1}),
the only anti-self-dual four-forms $F_1^{-}, F_2^{-}$ and $F_3^{-}$ 
are survived.

$\bullet$ Nonsupersymmetric $SO(7)^{+}$ critical point

One \cite{dN84} obtains the critical values
\bea
\la & = & \sqrt{2}\sinh^{-1}\left(\sqrt{\frac{1}{2}(\frac{3}{\sqrt{5}}-1)}\right)
=\sqrt{\la'^2+\rho^2}, \nonu \\ 
\alpha & = & 0=\frac{1}{2} \cos^{-1} \left(
\frac{\la'^2 \cos2\phi + \rho^2 \cos2\varphi}{\la'^2+\rho^2} \right),
\label{cond1}
\eea
after solving the four conditions characterized by    
(\ref{rel}), (\ref{rel1}) and the first and the fourth of (\ref{firstder}).  
Then it is clear to check the remaining second, third, fifth
and sixth equations of
(\ref{firstder}) satisfy under the condition 
$
\la' \sin \phi =0$.
This implies, under the condition for nonzero $\la'$(For $\la'=0$,
then we end up with the case discussed in section 2), that 
\bea
\phi =0 =\varphi,
\label{extra1}
\eea
by eliminating the $\rho$ field from two equations (\ref{cond1}).
Therefore, the additional field $\varphi$ is fixed as the $\phi$ field 
and the $\la'$ field is not fixed but the square sum $\la'^2 +\rho^2$
is fixed according to (\ref{cond1}).
This result is summarized in Table 1 where the five fields are fixed.
Since the angles $\alpha, \phi$ and $\varphi$ are equal to
zero via (\ref{cond1}) and (\ref{extra1}) and from (\ref{phiijkl1}),
the only self-dual four-forms $F_1^{+}, F_2^{+}$ and $F_3^{+}$ are survived.

$\bullet$ ${\cal N}=1$ $G_2$ critical point

One \cite{dWNW} has the critical values
\bea
\la  & = & \sqrt{2}\sinh^{-1}\left(\sqrt{\frac{2}{5}(\sqrt{3}-1)} \right)=
\sqrt{\la'^2+\rho^2}, \nonu \\
\alpha & = & \cos^{-1} \frac{\sqrt{3-\sqrt{3}}}{2}=\frac{1}{2} \cos^{-1} \left(
\frac{\la'^2 \cos2\phi + \rho^2 \cos2\varphi}{\la'^2+\rho^2} \right).
\label{cond2}
\eea
Then it is straightforward to check the remaining second, third, fifth
and sixth equations of
(\ref{firstder}) satisfy without any further constraints on the fields.
Therefore, the additional field $\varphi$ is not fixed and 
the $\la'$ field is not fixed but the square sum $\la'^2 +\rho^2$
is fixed according to (\ref{cond2}).

Let us introduce $F(\la',\phi,\varphi)$ defined by
\bea
F \equiv 
 \frac{\la'^2 \cos2\phi +
  \left(\left[\sqrt{2}\sinh^{-1}\sqrt{\frac{2}{5}(\sqrt{3}-1)} 
\right]^2 -\la'^2\right) \cos2\varphi}{
\left[\sqrt{2}\sinh^{-1}\sqrt{\frac{2}{5}(\sqrt{3}-1)}\right]^2}
-  \cos (2\cos^{-1} \frac{\sqrt{3-\sqrt{3}}}{2})
\label{surface}
\eea
by eliminating the $\rho$ field from (\ref{cond2})
and we present this three-dimensional contour plot  for
$F(\la',\phi,\varphi)=0$
in the Figure 1.
The maximal value of $\la'$ is given by 
\bea
\la'_{max}=\sqrt{2}\sinh^{-1}\sqrt{\frac{2}{5}
(\sqrt{3}-1)}=0.73
\label{lmax}
\eea 
when $\rho_{min}=0$ 
in (\ref{cond2}) and the corresponding $\phi$ is
given by $\phi=\cos^{-1} \frac{\sqrt{3-\sqrt{3}}}{2}=0.97$ by
substituting $\rho_{min}=0$ in the second equation of (\ref{cond2}). On the
other hand, the minimum value of $\la'$ is 
\bea
\la'_{min}=0
\label{lmin}
\eea 
when 
$\rho_{max}=\sqrt{2}\sinh^{-1}\sqrt{\frac{2}{5}
(\sqrt{3}-1)}=0.73$ through (\ref{cond2}) at which the
$\varphi$ is given by 
$\varphi=\cos^{-1} \frac{\sqrt{3-\sqrt{3}}}{2}=0.97$ similarly.
For $\la'_{min} < \la' < \la'_{max}$ with (\ref{lmax}) and
(\ref{lmin}) 
corresponding to nonzero
$\rho$ except the above two extreme limits,
any point on the surface can give the $G_2$ critical point.
The self-dual four-forms $F_1^{+}, F_2^{+}$ and $F_3^{+}$ 
and anti-self-dual four forms $F_1^{-}, F_2^{-}$ and $F_3^{-}$ 
are present.
This result is summarized in Table 1 where the three fields among six
fields are fixed.
Further constraints on these three remaining fields can be found
by requiring either the possibility of scalar potential in terms of
superpotential or the existence of BPS domain-wall solutions later.

\begin{figure}[ht]
   \epsfxsize=3.0in 
\centerline{\epsffile{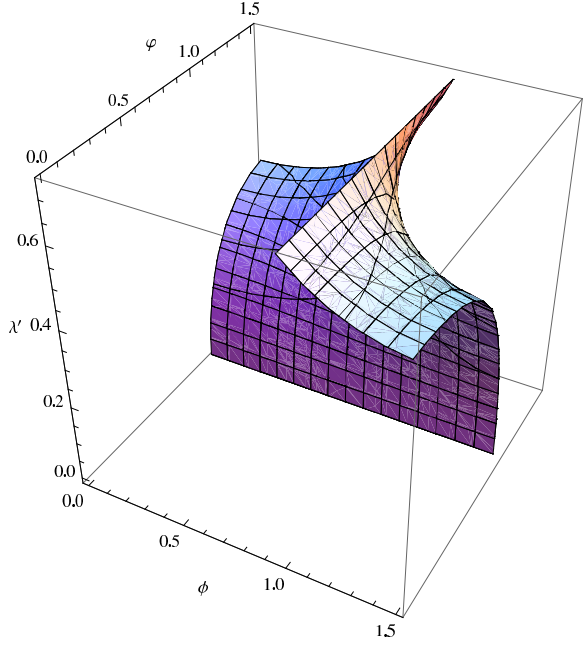}}
   \caption[FIG. \arabic{figure}.]{ 
\sl The three-dimensional contour plot for $F(\la',\phi,\varphi)=0$
defined by (\ref{surface}). The maximal value of $\la'$ is given by 
$\la'_{max}=\sqrt{2}\sinh^{-1}\sqrt{\frac{2}{5}
(\sqrt{3}-1)}=0.73$ when $\rho_{min}=0$ 
in (\ref{cond2}) and the corresponding $\phi$ is
given by $\phi=\cos^{-1} \frac{\sqrt{3-\sqrt{3}}}{2}=0.97$. On the
other hand, the minimum value of $\la'$ is $\la'_{min}=0$ when 
$\rho_{max}=\sqrt{2}\sinh^{-1}\sqrt{\frac{2}{5}
(\sqrt{3}-1)}=0.73$ at which the
$\varphi$ is given by 
$\varphi=\cos^{-1} \frac{\sqrt{3-\sqrt{3}}}{2}=0.97$ similarly.
For $\la'_{min} < \la' < \la'_{max}$ corresponding to nonzero
$\rho$,
any point on the surface can give the $G_2$ critical point.    }
\end{figure}

$\bullet$ Nonsupersymmetric $SU(4)^{-}$ critical point

One \cite{PW} gets the following critical values
\bea
\la=0, \qquad \sqrt{\la'^2+\rho^2}=\sqrt{2}\sinh^{-1}\left( 1 \right),
\qquad \frac{1}{2} \cos^{-1} \left(
\frac{\la'^2 \cos2\phi + \rho^2 \cos2\varphi}{\la'^2+\rho^2} \right)
=\frac{\pi}{2},
\label{cond3}
\eea
after solving the four conditions characterized by    
(\ref{rel}), (\ref{rel1}) and the first and the fourth of (\ref{firstder}).  
Then it is straightforward to check the remaining second, third, fifth
and sixth equations of
(\ref{firstder}) satisfy under the condition 
$
\la' \cos \phi =0$.
This implies, under the condition for nonzero $\la'$(For $\la'=0$,
then we end up with the case which was discussed in section 2), that 
\bea
\phi =\frac{\pi}{2} =\varphi,
\label{extra3}
\eea
by eliminating the $\rho$ field from two equations (\ref{cond3}).
Therefore, the additional field $\varphi$ is fixed as the $\phi$ field 
via (\ref{extra3}) and 
the $\la'$ field is not fixed but the square sum $\la'^2 +\rho^2$
is fixed according to (\ref{cond3}).
This result is summarized in Table 1 where the four fields among six
fields are fixed.
Since the angles $\phi$ and $\varphi$ are equal to
$\frac{\pi}{2}$ via (\ref{cond3})  and (\ref{extra3}) and 
$\la=0$,  from (\ref{phiijkl1}),
the only anti-self-dual four-forms $F_2^{-}$ and $F_3^{-}$ is survived.

$\bullet$ ${\cal N}=2$ $SU(3) \times U(1)$ critical point 

One \cite{NW} has the following values
\bea
\la & = & \sqrt{2}\sinh^{-1}\left(\frac{1}{\sqrt{3}} \right),
\quad  \sqrt{\la'^2+\rho^2} = \sqrt{2}\sinh^{-1}\left(\frac{1}{\sqrt{2}} \right),
\nonu \\ 
\alpha & = & 0, \qquad 
\frac{1}{2} \cos^{-1} \left(
\frac{\la'^2 \cos2\phi + \rho^2 \cos2\varphi}{\la'^2+\rho^2} \right)
=\frac{\pi}{2},
\label{cond4}
\eea
after solving the four conditions characterized by    
(\ref{rel}), (\ref{rel1}) and the first and the fourth of (\ref{firstder}).  
Then it is straightforward to check the remaining second, third, fifth
and sixth equations of
(\ref{firstder}) satisfy under the condition 
$
\la' \cos \phi =0$.
This implies, under the condition for nonzero $\la'$(For $\la'=0$,
then we end up with the case discussed in section 2), that 
\bea
\phi =\frac{\pi}{2} =\varphi,
\label{extra4}
\eea
by eliminating the $\rho$ field from two equations (\ref{cond4}).
Therefore, the additional field $\varphi$ is fixed as the $\phi$ field 
via (\ref{extra4}) and 
the $\la'$ field is not fixed but the square sum $\la'^2 +\rho^2$
is fixed according to (\ref{cond4}).
Since the angles $\phi$ and $\varphi$ are equal to
$\frac{\pi}{2}$ via (\ref{cond4})  and (\ref{extra4}), and 
$\alpha=0$, from (\ref{phiijkl1}),
the self-dual four-form $F_1^{+}$ and 
anti-self-dual four forms $F_2^{-}$ and $F_3^{-}$ are survived.

We summarize the analysis for the critical points in Table 1.

\bea
\begin{array}{|c|c|c|c|c|}
\hline $\mbox{Symmetry}$ & s, s', \al, \phi & \varphi & W=|z_3| & V \nonu \\
\hline
   SO(8) 
& s = 0 = s' & \mbox{many}
  & 1
 & -6 g^2 \nonu \\
\hline
   SO(7)^{-}  
& s=\pm \frac{1}{2}, s'=\frac{1}{2}, \al=\frac{\pi}{2} =\phi
  & \varphi=\frac{\pi}{2}
& \frac{3 \times 5^{3/4}}{8}  &- \frac{25\sqrt{5}}{8}g^2  \nonu \\
\hline
   SO(7)^{+}    & 
 s=\sqrt{\frac{1}{2}(\frac{3}{\sqrt{5}}-1)} =s', 
 \al=0=\phi & \varphi=0
 & \frac{3}{2} \times 5^{-1/8} & - 2\times5^{3/4}g^2 \nonu \\
\hline
   G_{2}   
&  s =\pm \sqrt{\frac{2}{5}(\sqrt{3}-1)}, s'=\sqrt{\frac{2}{5}(\sqrt{3}-1)},
 & &  & \nonu \\
& \al = \cos^{-1} \frac{\sqrt{3-\sqrt{3}}}{2} =\beta 
 &\mbox{many} & \sqrt{ \frac{36\sqrt{2} \times 3^{1/4}}{25\sqrt{5}}} 
& - \frac{216\sqrt{2} \times 3^{1/4}}{25\sqrt{5}}g^2
\nonu \\ \hline
   SU(4)^{-}    
&  s=0, s'=1, \phi=\frac{\pi}{2}   & \varphi=\frac{\pi}{2} & \frac{3}{2}  & 
- 8g^2 \nonu \\ \hline
 SU(3) \times U(1)    
&  s=\frac{1}{\sqrt{3}}, s'=\frac{1}{\sqrt{2}}, \al=0,
\phi=\frac{\pi}{2}& \varphi=\frac{\pi}{2} 
& \frac{3^{3/4}}{2} & - \frac{9 \sqrt{3}}{2} g^2 \nonu
\\ \hline
\end{array}
\nonu
\eea
Table 1. \sl Symmetry group,
vacuum expectation values of fields, superpotential and 
cosmological constants.  The $s, s'$  and $\beta$ are defined as (\ref{csrelation}). 
Compared to the section 2 parametrized by
four fields,
the main difference arises as the fact that these critical points except $G_2$
symmetric case  have $\varphi=\phi$ which is fixed while for $G_2$ symmetric case, 
the fields $\phi$ and $\varphi$ are not fixed but 
constrained to the surface equations (\ref{cond2}).
The field $\la'$ is replaced by $\sqrt{\la'^2 + \rho^2}$ which is common to
all of these. These vacuum expectation values are only obtained from the
criticality of scalar potential.
\rm

Are these critical points for the scalar potential those for the
superpotential also? The superpotential is given by 
\bea 
W(\la, \la',\rho; \al, \phi, \varphi) & = & |z_3|, \nonu \\
z_3(\la, \la',\rho; \al, \phi, \varphi) & = & 6e^{i(\alpha + 2\beta)}p^2qr^2t^2
 + 6e^{2i(\alpha + \beta)}pq^2r^2t^2  + p^3(r^4 + e^{4i\beta}t^4)  \nonu \\
& & +
e^{3i\alpha}q^3(r^4 + e^{4i\beta}t^4),
\label{Super}
\eea
with (\ref{pqrt1}) and (\ref{csrelation}). One can take the
superpotential as $z_2$ via (\ref{Relation}).
Then it is obvious that the conditions for 
the critical points of the superpotential
are given by 
\bea
\frac{\pa W}{\pa \la}  = \frac{\pa W}{\pa \la'} =
\frac{\pa W}{\pa \rho} =
\frac{\pa W}{\pa \alpha}  = 
\frac{\pa W}{\pa \phi} =
\frac{\pa W}{\pa \varphi} =0, 
\label{Wcond}
\eea
and we have checked that
the above $G_2$ critical point for the scalar
potential are also 
the critical points for the superpotential satisfying (\ref{Wcond}). 
For other critical points
except the $SO(8)$ case, they are not critical points of superpotential.   
Note that when we compute 
$\frac{\pa W}{\pa \rho}$ or $\frac{\pa W}{\pa \la'}$,
the denominator contains 
\bea
\sin^2 2\beta=\sqrt{1-\left(\frac{\la'^2 \cos2\phi + \rho^2 \cos2\varphi}
{\la'^2+\rho^2}\right)^2}
\label{den}
\eea
with (\ref{csrelation}).
Then at the $SO(7)^{-}, SO(7)^{+}, SU(4)^{-}$ and $SU(3) \times U(1)$
critical points, 
the above (\ref{den}) vanishes because the values $\phi=\varphi$ are
either zero or $\frac{\pi}{2}$. This implies that the derivatives of
$W$ with respect to $\rho$ or $\la'$ at these critical points are infinite.
This is rather different feature from the previous results in \cite{AW}.
This is due to the fact that the above $\beta$ is a function of 
$\la',\rho,\phi$ and $\varphi$ and when  
the derivatives of $W$ with respect to the fields $\la'$ or $\rho$
are performed, the above phenomenum corresponding to (\ref{den}) occurs. 
Although we have not found the scalar potential in terms of a
superpotential for generic six fields explicitly at the moment, 
we know that the
derivative terms of $W$ with respect to the six fields at the $G_2$
critical
point vanish(recall that this critical condition does not provide the
infinity of the derivative $W$ with respect to $\la'$ or $\rho$,
contrary to the feature in (\ref{den}) because (\ref{den}) is not
equal to zero. See the Table 1.) 
and there exists a relation $V=-6g^2 W^2$ at the
critical points. 
This implies that if the scalar potential is made of  
derivative terms of $W$ with respect to the six fields(together with
very complicated coefficient functions which depend on them) and 
square of $W$,
the supersymmetry preserving vacua have negative cosmological constant
and the critical points of $W$ yield supersymmetric stable $AdS_4$
vacua in gauged ${\cal N}=8$ supergravity.

Are there any BPS domain-wall solutions?
As long as the field $\varphi$ approaches the field $\phi$, a simple
relation between the scalar potential and superpotential arises.
When 
\bea
\varphi(r)=\phi(r),
\label{varphiphi}
\eea  
from (\ref{csrelation}), one obtains 
$\varphi=\phi=\beta$ and the scalar potential and its superpotential
depend on $\la, \sqrt{\la'^2 +\rho^2}, \alpha$ and $\phi$. 
Of course, the above critical points satisfy this condition and 
in particular, the $G_2$ critical point has other solutions for 
$\varphi \neq \phi$(See the Figure 1).
Then it is straightforward, from the experience of previous section, to check
that the scalar potential, under the condition (\ref{varphiphi}), 
is 
\bea 
V(\la, \la',\rho, \al, \phi, \varphi)|_{\varphi=\phi}= 
g^2 \left[ \frac{16}{3} \left|\frac{\partial
z_3}{\partial \la} \right|^2 +  \left|\frac{\sqrt{\la'^2 +\rho^2}}{
\la'} \frac{\partial z_3}{\partial
\la'}+  \frac{\sqrt{\la'^2 +\rho^2}}{
\rho} \frac{\partial z_3}{\partial \rho}\right|^2 - 6  \left|z_3\right|^2 \right]
\label{reducedpot}
\eea
where $z_3$ is given by (\ref{superpotential2}) or (\ref{Super})
together with the constraint (\ref{varphiphi}).
From the kinetic terms (\ref{kinetic2}), one realizes that 
one recombines the first four kinetic terms of (\ref{simplekin})
with the corresponding scalar potential parts (\ref{reducedpot}) in terms of the squares
of the four derivatives of superpotential with some algebraic
relations. 
Then we are
left with the last term of (\ref{simplekin})  
which can appear in the integrand of functional.    
In order to have the
BPS bound, inequality of the energy-density, one should have
$\la'\pa_{\mu} \rho
=\rho \pa_{\mu} \la'$. Then it is easy to
check that the solution provides 
\bea
\la'(r) =\rho(r).
\label{larho}
\eea  
Therefore, one should take $\la'(r)=\rho(r)$ along the
whole RG flow.  
Then one can check the following scalar potential 
\bea 
V(\la, \la',\rho, \al, \phi, \varphi)|_{\varphi=\phi,\rho=\la'}= 
g^2 \left[ \frac{16}{3} \left|\frac{\partial
z_3}{\partial \la} \right|^2 + 2 \left| \frac{\partial z_3}{\partial
\la'}\right|^2 - 6  \left|z_3\right|^2 \right],
\label{reducedpotential}
\eea
where $z_3$ is given by (\ref{superpotential2}) or (\ref{Super})
together with the constraints (\ref{varphiphi}) and (\ref{larho}).
This can be done by substituting the conditions (\ref{varphiphi}) and
(\ref{larho}) 
into the
scalar potential and superpotential and then 
looking for the relation between them, in order to get the correct
coefficient 2 in (\ref{reducedpotential}), rather than 
by using (\ref{reducedpot}) directly.
By analyzing the energy-density from the kinetic terms (\ref{simplekin}) and scalar
potential
(\ref{reducedpotential}), as done in \cite{AW}, one arrives at 
the following first order differential equations 
\bea  
\frac{d \la}{d r} & = & \pm \frac{8\sqrt{2}}{3}g \partial_{\la} W
,\nonu \\
\frac{d \la'}{d r} & = & \pm \sqrt{2} g \pa_{\la'} W = \frac{d \rho}{d r}, \nonu \\
\frac{d \alpha}{ d r} & = & \pm
\frac{\sqrt{2}}{3p^2 q^2} g \pa_{\al} W, \nonu \\
\frac{d \phi}{d r}   & = &  \pm \frac{\sqrt{2}}{4r^2 t^2} g \pa_{\phi} W
=\frac{d \varphi}{d r}, \nonu \\ 
\frac{d A}{ d r} & = & \mp \sqrt{2} g W,
\label{BPS}
\eea
where $W=|z_3|$ as in (\ref{Super}) with the constraints (\ref{varphiphi}) and
(\ref{larho}), the relations (\ref{pqrt1}) with these constraints are used  and
the scale factor $A(r)$ in the last equation 
appears in the four-dimensional metric
$
ds^2= e^{2A(r)} \eta_{\mu \nu} dx^{\mu} dx^{\nu} + dr^2$ with
three-dimensional metric $\eta_{\mu \nu}=(-,+,+)$. The superpotential
$W$ appearing in (\ref{BPS}) has 
almost the same as the one in \cite{AW} except the factor
$\sqrt{2}$ in front of $\la'$(or $\rho$) for the hyperbolic functions of 
(\ref{pqrt1}). 
This is the reason why the second
equation of (\ref{BPS}) is different from those of \cite{AW} and
moreover, there are extra two first order differential equations 
on $\rho$ and $\varphi$.
Note that these BPS equations hold for the restricted submanifold by (\ref{varphiphi})
and (\ref{larho}) when we have $SU(3)$-singlet space characterized 
by (\ref{phiijkl1}). 

The deformation analysis in three-dimensional boundary field theory 
can be done by recalling that there exist generalized four-forms 
given in (\ref{four}) and adding the right fermion mass terms as we
did in \cite{Ahn0806n1}. We expect to have the additional terms in the
fermionic mass term due to 
the presence of $F_3^{\pm}$.
Then by requiring the appropriate number of supersymmetry 
one can impose some constraints on the supersymmetry parameter.
By looking at the bosonic and fermionic mass terms, one can construct the variation
of the bosonic mass term and fermionic mass term. 
Then the final bosonic mass term can be determined by vanishing of
this variation of Lagrangian. 
In ${\cal N}=1$ superspace notation, the deformed superpotential looks like
\bea
\Delta W = \frac{1}{2} m_7 \Tr \Phi_7^2 +\frac{1}{2} m_8 \Tr \Phi_8^2
+\frac{1}{2} m_{78} \Tr \Phi_7 \Phi_8.
\label{deformedW}
\eea
We  turn on the mass perturbation in the UV and flow
to the IR. This maps to turning on certain fields in 
the $AdS_4$ supergravity where they approach to zero in the UV and
develop a nontrivial profile as a function of $r$ as one goes to the IR. 
Note that from (\ref{S1}), the $(78)$- and $(87)$-component are given
by $\rho \cos \varphi$. For nonzero $\rho$ and nonzero $\cos \varphi$, 
this provides the mass term of cross terms for (\ref{deformedW}).
For the ${\cal N}=2$ $SU(3) \times U(1)$ supersymmetric flow($m_7,m_8 \neq 0 $)  
in which $\varphi
=\frac{\pi}{2}$ at the IR fixed point 
from the Table 1, the boson mass matrix can be
diagonalized.
For the ${\cal N}=1$ $G_2$ supersymmetric flow($m_7=0$ or $m_8=0$)  
in which $\varphi$ can have any
values on the constrained surface (\ref{surface})(even if
$\varphi=\phi$ in (\ref{varphiphi}), 
this value is not equal to $\frac{\pi}{2}$ but $0.97$) 
at the IR fixed point 
from the Table 1, the degenerate vacua imply the above mass term
of cross terms of (\ref{deformedW}). 
Therefore, the additional supergravity fields $(\rho, \varphi)$
in this section 
correspond to and are dual to the mass term $m_{78}$ of cross terms in the ${\cal
N}=1$ or ${\cal N}=2$ superconformal Chern-Simons matter theory.  
See the reference \cite{Ahn0905} for more details.

\section{Conclusions and outlook}


By considering the most general $SU(3)$ singlet space of
gauged ${\cal N}=8$ supergravity,
we have found the new scalar potential (\ref{Poten}). 
The critical points of the scalar potential are summarized in Table 1.
Although the critical values of the scalar potential are the same as
the ones in section 2, one, two, or three 
of six fields is not determined completely.  
In other words, there are infinite solutions to provide the same
cosmological constant.
In particular, for the ${\cal N}=1$ $G_2$ critical point, 
the constraint surface parametrized by three scalar fields(i.e., three
fields are fixed) on which 
the cosmological constant has same value is drawn in Figure 1.
The BPS domain-wall solutions for restricted scalar
submanifold are presented in (\ref{BPS}).
The three-dimensional mass-deformed 
superconformal Chern-Simons matter theory is discussed in 
(\ref{deformedW}).

$\bullet$ An eleven-dimensional lift

It is interesting to uplift the four-dimensional gauged supergravity
we described here to eleven-dimensions. This has been done in \cite{CPW}
or \cite{AI}. According to \cite{dWNW}, one can construct the
eleven-dimensional metric from the metric of the solutions to
four-dimensional gauged ${\cal N}=8$ supergravity.
The nontrivial task is how to find out the correct expression for the
internal four-form flux which will be present for the 
the most general $SU(3)$ singlet space of
gauged ${\cal N}=8$ supergravity we described here.

$\bullet$ The applications of this $SU(3)$ invariant sector

In \cite{AW02}, the $G_2$ sector of noncompact $SO(7,1)$ gauging, the 
$SU(3)$ sector of noncompact $SO(6,2)$ gauging, 
the $G_2$ sector of nonsemisimple Inonu-Wigner(IW) 
contraction $CSO(7,1)$ gauging,
and the 
$SU(3)$ sector of nonsemisimple  IW contraction $CSO(6,2)$ gauging 
are studied. 
It would be interesting to find whether there exist any new critical
points
of the scalar potential restricted to the $SU(3)$-singlet sector we
studied in this paper for 
the $SO(p, 8-p)$ and $CSO(p,8-p)$ gaugings.

$\bullet$ Any BPS equations for arbitrary $\la,
\la',\rho,\alpha,\phi,\varphi$

In this paper, we have only considered the BPS equations (\ref{BPS}) for
$\varphi=\phi, \la'=\rho$
in which the kinetic terms are simple and 
the scalar potential can be written in terms of a superpotential.
It is natural to ask whether there are any BPS equations for general vacuum
expectation values or not. In \cite{BHPW}, for the kinetic terms, 
three complex fields are introduced. Moreover, the supergravity
potential on the $SU(3)$ invariant sector was given in terms of the
derivatives
of superpotential with respect to these complex fields.   
The first step in this direction is to find out the simplest kinetic
terms by using the change of variables and the next step is to rewrite
a scalar potential in terms of a superpotential for arbitrary vacuum
expectation values. 

$\bullet$ Geometric superpotential

One can go to the $SL(8,{\bf R})$ basis \cite{KW} and  rotate the 
28-beins (\ref{uv1}), (\ref{u1}), (\ref{Change}), (\ref{v1}), and
(\ref{Change1}) 
using $SO(8)$ generators. Then
geometric T-tensor can be written in terms of rectangular coordinates
in ${\bf R}^8$. Then it is straightforward to construct the
corresponding $A_1$ tensor and its geometric superpotential can be
read off. This will provide some hints for the eleven-dimensional lift
for the metric and solutions. 

$\bullet$ The RG flow connecting  ${\cal N}=1$ $G_2$ fixed point 
to ${\cal N}=2$ $SU(3) \times U(1)$ fixed point 

As mentioned in the introduction, the results of \cite{BHPW} 
show that there exists an ${\cal N}=1$ supersymmetric flow from the 
$G_2$ symmetric point to the $SU(3) \times U(1)$ symmetric point under
the parametrization of section 2. So it is natural to ask 
what happens for the RG flow connecting  ${\cal N}=1$ $G_2$ fixed point 
to ${\cal N}=2$ $SU(3) \times U(1)$ fixed point under the
parametrization of section 3. It would be interesting to find the
right superpotential depending on only three variables $(\la, \la',\rho)$  
and the other angle variables $(\alpha, \phi, \varphi)$ 
should be written in terms of $\la'$ and $\rho$.  

\vspace{.7cm}

\centerline{\bf Acknowledgments}

This work was supported by the 
National Research Foundation of Korea(NRF) grant 
funded by the Korea government(MEST)(No. 2009-0084601).

\appendix

\renewcommand{\thesection}{\large \bf \mbox{Appendix~}\Alph{section}}
\renewcommand{\theequation}{\Alph{section}\mbox{.}\arabic{equation}}
\section{The $28 \times 28$ matrices $u$ and $v$, $A_2$ tensor,
  kinetic terms of section 2}

The 28-beins $u^{IJ}_{\;\;\;KL}$ and $v^{IJKL}$ 
fields, which are elements of $56 \times 56$ ${\cal V}(x)$ of
the fundamental 56-dimensional representation of $E_{7(7)}$ through (\ref{56bein}),
can be obtained 
by exponentiating the vacuum expectation values $\phi_{ijkl}$ 
(\ref{para}) via (\ref{calV}). These 28-beins have the
following single $4 \times 4$ block diagonal matrix $u_1$ and $v_1$
and three $8 \times 8$ block diagonal matrices $u_i$ and $v_i$ where
$i=2, 3, 4$ respectively:
\bea 
u^{IJ}_{\;\;\;KL}  =  \mbox{diag}
(u_1,u_2,u_3,u_4), \qquad
v^{IJKL}  = 
\mbox{diag} (v_1,v_2,v_3,v_4). 
\label{uv}
\eea 
Each hermitian(for example, 
$(u_1)^{78}_{\;\;\;12}=((u_1)^{12}_{\;\;\;78})^{\ast}=BK^2$) 
submatrix is $4 \times 4$ matrix or $8 \times 8$ matrix and we denote
antisymmetric index pairs $[IJ]$ and $[KL]$ explicitly for convenience.
For simplicity, we make an empty space corresponding to lower triangle 
elements that can be read off from the corresponding upper triangle
elements by hermiticity. Then the $4 \times 4$ submatrix and $8\times
8$ submatrices lead to the following expression
\bea 
u_{1}  =  \left(\begin{array}{ccccc}
 & [12] & [34] & [56] & [78]\\
{}[12] & A & B & B & K^{-2}B\\
{}[34] &  & A & B & K^{-2}B\\
{}[56] &  &  & A & K^{-2}B\\
{}[78] &  &  &  & A\end{array}\right),\nonu 
\eea
\bea
u_{2}  =  \left(\begin{array}{ccccccccc}
 & [13] & [24] & [57] & [68] & [14] & [23] & [58] & [67]\\
{}[13] & C & -D & -E K^{-1} & E^{*} K^{-1} & 0 & 0 & -FK^{-1} & -F^{*}K^{-1}\\
{}[24] &  & C & EK^{-1} & -E^{*}K^{-1} & 0 & 0 & FK^{-1} & F^{*}K^{-1}\\
{}[57] &  &  & C & -DG^{*} & -KF & -KF & 0 & -DH\\
{}[68] &  &  &  & C & KF^{*} & KF^{*} & -DH & 0\\
{}[14] &  &  &  &  & C & D & E^{*}K^{-1} & EK^{-1}\\
{}[23] &  &  &  &  &  & C & E^{*}K^{-1} & EK^{-1}\\
{}[58] &  &  &  &  &  &  & C & DG\\
{}[67] &  &  &  &  &  &  &  & C\end{array}\right),
\nonu 
\eea
\bea
u_{3}  =  \left(\begin{array}{ccccccccc}
 & [15] & [26] & [37] & [48] & [16] & [25] & [38] & [47]\\
{}[15] & C & -D & EK^{-1} & -E^{*}K^{-1} & 0 & 0 & FK^{-1} & F^{*}K^{-1}\\
{}[26] &  & C & -EK^{-1} & -E^{*}K^{-1} & 0 & 0 & -FK^{-1} & -F^{*}K^{-1}\\
{}[37] &  &  & C & -DG^{*} & KF & KF & 0 & -DH\\
{}[48] &  &  &  & C & -KF^{*} & -KF^{*} & -DH & 0\\
{}[16] &  &  &  &  & C & D & -E^{*}K^{-1} & -EK^{-1}\\
{}[25] &  &  &  &  &  & C & -E^{*}K^{-1} & -EK^{-1}\\
{}[38] &  &  &  &  &  &  & C & DG\\
{}[47] &  &  &  &  &  &  &  & C\end{array}\right),
\nonu \eea
\bea
u_{4}  =  \left(\begin{array}{ccccccccc}
 & [17] & [28] & [35] & [46] & [18] & [27] & [36] & [45]\\
{}[17] & C & -DG^{*} & -KE^{*} & KE^{*} & 0 & -DH & -KF & -KF\\
{}[28] &  & C & KE & -KE & -DH & 0 & KF^{*} & KF^{*}\\
{}[35] &  &  & C & -D & -FK^{-1} & -F^{*}K^{-1} & 0 & 0\\
{}[46] &  &  &  & C & FK^{-1} & F^{*}K^{-1} & 0 & 0\\
{}[18] &  &  &  &  & C & DG & KE & KE\\
{}[27] &  &  &  &  &  & C & KE^{*} & KE^{*}\\
{}[36] &  &  &  &  &  &  & C & D\\
{}[45] &  &  &  &  &  &  &  & C\end{array}\right),
\label{u}
\eea 
where
we introduce some quantities that are functions of $\la, \la', \alpha,
\phi, \theta$, and $\psi$ with (\ref{pqrt})
as follows:
\bea
&& A \equiv p^3,\qquad B \equiv pq^2,\qquad
C \equiv pr^2,\qquad 
D \equiv pt^2, \nonu \\
&& E \equiv \left[\cos\phi\cos(\theta+\psi)+i\sin\phi\cos(\theta-\psi) 
\right] q r t, \nonu \\
&& F \equiv \left[\cos\phi\sin
(\theta+\psi)+i\sin\phi\sin(\theta-\psi) \right] q r t, \nonu \\
&&
G \equiv (\cos 2\phi+i\cos2\theta\sin2\phi),
\qquad
H \equiv i \sin2\theta\sin2\phi, \qquad
K \equiv e^{i\alpha},
\label{change}
\eea
and 
\bea
v_{1}  =  \left(\begin{array}{ccccc}
 & [12] & [34] & [56] & [78]\\
{}[12] & K^{-1}L & K^{-1}M & K^{-1}M & KM\\
{}[34] &  & K^{-1}L & K^{-1}M & KM\\
{}[56] &  &  & K^{-1}L & KM\\
{}[78] &  &  &  & K^{3}L\end{array}\right),
\nonu 
\eea
\bea
v_{2}  =  \left(\begin{array}{ccccccccc}
 & [13] & [24] & [57] & [68] & [14] & [23] & [58] & [67]\\
{}[13] & JK^{-1} & -NK^{-1} & -P^{*} & P & 0 & 0 & -Q^{*} & -Q\\
{}[24] &  & JK^{-1} & P^{*} & -P & 0 & 0 & Q^{*} & Q\\
{}[57] &  &  & KJR^{*} & -KN & -Q & -Q & -KJS & 0\\
{}[68] &  &  &  & KJR & Q^{*} & Q^{*} & 0 & -KJS\\
{}[14] &  &  &  &  & JK^{-1} & NK^{-1} & P & P^{*}\\
{}[23] &  &  &  &  &  & JK^{-1} & P & P^{*}\\
{}[58] &  &  &  &  &  &  & KJR & KN\\
{}[67] &  &  &  &  &  &  &  & KJR^{*}\end{array}\right),\nonu 
\eea
\bea
v_{3}  =  \left(\begin{array}{ccccccccc}
 & [15] & [26] & [37] & [48] & [16] & [25] & [38] & [47]\\
{}[15] & JK^{-1} & -NK^{-1} & P^{*} & -P & 0 & 0 & Q^{*} & Q\\
{}[26] &  & JK^{-1} & -P^{*} & P & 0 & 0 & -Q^{*} & -Q\\
{}[37] &  &  & KJR^{*} & -KN & Q & Q & -KJS & 0\\
{}[48] &  &  &  & KJR & -Q^{*} & -Q^{*} & 0 & -KJS\\
{}[16] &  &  &  &  & JK^{-1} & NK^{-1} & -P & -P^{*}\\
{}[25] &  &  &  &  &  & JK^{-1} & -P & -P^{*}\\
{}[38] &  &  &  &  &  &  & KJR & KN\\
{}[47] &  &  &  &  &  &  &  & KJR^{*}\end{array}\right),\nonu 
\eea
\bea
v_{4}  =  \left(\begin{array}{ccccccccc}
 & [17] & [28] & [35] & [46] & [18] & [27] & [36] & [45]\\
{}[17] & KJR^{*} & -KN & -P^{*} & P^{*} & -KJS & 0 & -Q & -Q\\
{}[28] &  & KJR & P & -P & 0 & -KJS & Q^{*} & Q^{*}\\
{}[35] &  &  & JK^{-1} & -NK^{-1} & -Q^{*} & -Q & 0 & 0\\
{}[46] &  &  &  & JK^{-1} & Q^{*} & Q & 0 & 0\\
{}[18] &  &  &  &  & KJR & KN & P & P\\
{}[27] &  &  &  &  &  & KJR^{*} & P^{*} & P^{*}\\
{}[36] &  &  &  &  &  &  & JK^{-1} & NK^{-1} \\
{}[45] &  &  &  &  &  &  &  & JK^{-1} \end{array}\right),
\label{v}
\eea
where
some quantities that are also functions of $\la, \la',\alpha,\phi,\theta$ and $\psi$
with (\ref{pqrt}) are given by
\bea
&& J \equiv q t^2, \qquad L\equiv q^3,\qquad M \equiv p^2 q,\qquad
N \equiv q r^2, \nonu \\
&& P \equiv \left[\cos\phi\cos
(\theta+\psi)+i\sin\phi\cos
(\theta-\psi)\right] p r t, \nonu \\
&&
Q \equiv \left[\cos\phi\sin(\theta+\psi)+
i\sin\phi\sin(\theta-\psi)\right] p r t, \nonu \\
&&
R \equiv (\cos2\phi+i\cos2\theta\sin2\phi),
\qquad S \equiv i \sin2\theta\sin2\phi.
\label{change1}
\eea
The lower triangle part of (\ref{u}) and (\ref{v}) 
can be read off from the upper triangle part by
hermitian property.
Also, the other kinds of 
28-beins $u_{IJ}^{\;\;\;KL}$ and $v_{IJKL}$ are obtained by taking a
complex conjugation of (\ref{u}) and (\ref{v}). The complex conjugation operation 
can be done by raising or lowering the indices:  
$(u_{IJ}^{\;\;\;KL})^{\ast} =  u^{IJ}_{\;\;\;KL}$ and etc using 
(\ref{uv}), (\ref{u}), (\ref{v}), (\ref{change}) and (\ref{change1}).
The above $u_1$ and $v_1$ were obtained from the computation the $8
\times 8$ block diagonal matrix that is given by 
(\ref{zero}), (\ref{4x4}), (\ref{4x4new}), and (\ref{anotherzero}) 
explained in section 2. That is,
$u_1$ is two-two block diagonal while $v_1$ is two-one block diagonal. 
Of course, the complex conjugations of these appear as 
one-one block diagonal and one-two block diagonal respectively.

The final expression of kinetic terms, 
by counting the correct
multiplicities,  
$\frac{1}{96}|A_{\mu}^{\; ijkl}|^2 $ is given by
\bea
& & \frac{3}{8}(\pa_{\mu}\la)^{2}+
3p^2 q^2 (\pa_{\mu}\alpha)^{2}+\frac{1}{2}(\pa_{\mu}\la')^{2}+
4r^{2}t^{2}\left[(\pa_{\mu}\phi)^{2}+(1+2t^2)^2 (\pa_{\mu}\psi
)^{2}
  +  (1+2r^{2}t^{2})  (\pa_{\mu}\theta)^{2} \right] \nonu \\
&  & +
8r^2 (\pa_{\mu}\theta)
\left[(t+2t^{3})^{2}\cos 2\phi \pa_{\mu}\psi+
r^{2}t^{4}\cos 4\phi \pa_{\mu}\theta \right].
\label{KinKin}
\eea
This (\ref{KinKin}) can be rewritten as
\bea
\frac{1}{96}|A_{\mu}^{\; ijkl}|^2 & = & \frac{3}{8}(\pa_{\mu}\la)^{2}+3p^{2}
q^{2}(\pa_{\mu}\alpha)^{2}+\frac{1}{2}(\pa_{\mu}\la')^{2}+
4r^{2}t^{2}(\pa_{\mu}\phi)^{2} +a_{+}G_{+}^{2}+a_{-}G_{-}^{2},
\nonu
\eea
where
\bea
a_{\pm} & \equiv & \frac{(5+3\cosh 4\la')\sinh^2(2\la')+
2\cos4\phi \sinh^4(2\la') \pm2R_{\la'\phi}}{
16R_{\la'\phi}[R_{\la'\phi}\pm2\sin^2(2\phi)\sinh^4(2\la')]},
\nonu \\
G_{\pm} & \equiv & [R_{\la'\phi}\pm2\sin^2(2\phi)
\sinh^4(2\la')]\pa_{\mu}\psi \pm\cos(2\phi)\sinh^2(4\la') \pa_{\mu}\theta,
\nonu \\
R_{\la'\phi} & \equiv & \sqrt{4\sin^4(2\phi)
\sinh^8(2\la')+\cos^2(2\phi)\sinh^4(4\la')}.
\label{kinother}
\eea

\section{The $28 \times 28$ matrices $u$ and $v$, $A_2$ tensor,
  kinetic terms of section 3}

The 28-beins $u^{IJ}_{\;\;\;KL}$ and $v^{IJKL}$ 
fields, which are elements of $56 \times 56$ ${\cal V}(x)$ of
the fundamental 56-dimensional representation of $E_{7(7)}$ through (\ref{56bein}),
can be obtained 
by exponentiating the vacuum expectation values $\phi_{ijkl}$ 
(\ref{phiijkl1}) via (\ref{calV}). These 28-beins have the
following single $4 \times 4$ block diagonal matrix $u_1$ and $v_1$
and three $8 \times 8$ block diagonal matrices $u_i$ and $v_i$ where
$i=2, 3, 4$ respectively:
\bea 
u^{IJ}_{\;\;\;KL}  =  \mbox{diag}
(u_1,u_2,u_3,u_4), \qquad
v^{IJKL}  = 
\mbox{diag} (v_1,v_2,v_3,v_4). 
\label{uv1}
\eea 
Each hermitian
submatrix is $4 \times 4$ or $8 \times 8$ matrices and we denote
antisymmetric index pairs $[IJ]$ and $[KL]$ explicitly for convenience.
Then the $4 \times 4$ submatrix and $8\times
8$ submatrices lead to the following expression 
\bea
u_{1}  =  \left(\begin{array}{ccccc}
 & [12] & [34] & [56] & [78]\\{}
[12] & A & B & B & \frac{B}{K_{1}^{2}}\\{}
[34] &  & A & B & \frac{B}{K_{1}^{2}}\\{}
[56] &  &  & A & \frac{B}{K_{1}^{2}}\\{}
[78] &  &  &  & A\end{array}\right),\nonu 
\eea
\bea
u_{2}  =  \left(\begin{array}{ccccccccc}
 & [13] & [24] & [57] & [68] & [14] & [23] & [58] & [67]\\{}
[13] & C & -D & -\frac{K_{2}E}{K_{1}G} & \frac{E}{K_{1}K_{2}G} & 0 & 0
 & 
-\frac{F}{K_{1}K_{3}G} & -\frac{K_{3}F}{K_{1}G}\\{}
[24] &  & C & \frac{K_{2}E}{K_{1}G} & -\frac{E}{K_{1}K_{2}G} & 0 & 0 & 
\frac{F}{K_{1}K_{3}G} & \frac{K_{3}F}{K_{1}G}\\{}
[57] &  &  & C & -D_{1} & -\frac{K_{1}F}{K_{3}G} & -\frac{K_{1}F}{K_{3}G} & 0 & D_{3}\\{}
[68] &  &  &  & C & \frac{K_{1}K_{3}F}{G} & \frac{K_{1}K_{3}F}{G} & D_{3} & 0\\{}
[14] &  &  &  &  & C & D & \frac{E}{K_{1}K_{2}G} & \frac{K_{2}E}{K_{1}G}\\{}
[23] &  &  &  &  &  & C & \frac{E}{K_{1}K_{2}G} & \frac{K_{2}E}{K_{1}G}\\{}
[58] &  &  &  &  &  &  & C & D_{2}\\{}
[67] &  &  &  &  &  &  &  & C\end{array}\right),
\nonu 
\eea
\bea
u_{3}  =  \left(\begin{array}{ccccccccc}
 & [15] & [26] & [37] & [48] & [16] & [25] & [38] & [47]\\{}
[15] & C & -D & \frac{K_{2}E}{K_{1}G} & -\frac{E}{K_{1}K_{2}G} & 0 & 0
 & 
\frac{F}{K_{1}K_{3}G} & \frac{K_{3}F}{K_{1}G}\\{}
[26] &  & C & -\frac{K_{2}E}{K_{1}G} & \frac{E}{K_{1}K_{2}G} & 0 & 0 &
 -
\frac{F}{K_{1}K_{3}G} & -\frac{K_{3}F}{K_{1}G}\\{}
[37] &  &  & C & -D_{1} & \frac{K_{1}F}{K_{3}G} & \frac{K_{1}F}{K_{3}G} & 0 & D_{3}\\{}
[48] &  &  &  & C & -\frac{K_{1}K_{3}F}{G} & -\frac{K_{1}K_{3}F}{G} & D_{3} & 0\\{}
[16] &  &  &  &  & C & D & -\frac{E}{K_{1}K_{2}G} & -\frac{K_{2}E}{K_{1}G}\\{}
[25] &  &  &  &  &  & C & -\frac{E}{K_{1}K_{2}G} & -\frac{K_{2}E}{K_{1}G}\\{}
[38] &  &  &  &  &  &  & C & D_{2}\\{}
[47] &  &  &  &  &  &  &  & C\end{array}\right),
\nonu \eea
\bea
u_{4}  =  \left(\begin{array}{ccccccccc}
 & [17] & [28] & [35] & [46] & [18] & [27] & [36] & [45]\\{}
[17] & C & -D_{1} & -\frac{K_{1}E}{K_{2}G} & \frac{K_{1}E}{K_{2}G} & 0
 & D_{3} 
& -\frac{K_{1}F}{K_{3}G} & -\frac{K_{1}F}{K_{3}G}\\{}
[28] &  & C & \frac{K_{1}K_{2}E}{G} & -\frac{K_{1}K_{2}E}{G} & D_{3} &
 0 & 
\frac{K_{1}K_{3}F}{G} & \frac{K_{1}K_{3}F}{G}\\{}
[35] &  &  & C & -D & -\frac{F}{K_{1}K_{3}G} & -\frac{K_{3}F}{K_{1}G} & 0 & 0\\{}
[46] &  &  &  & C & \frac{F}{K_{1}K_{3}G} & \frac{K_{3}F}{K_{1}G} & 0 & 0\\{}
[18] &  &  &  &  & C & D_{2} & \frac{K_{1}K_{2}E}{G} & \frac{K_{1}K_{2}E}{G}\\{}
[27] &  &  &  &  &  & C & \frac{K_{1}E}{K_{2}G} & \frac{K_{1}E}{K_{2}G}\\{}
[36] &  &  &  &  &  &  & C & D\\{}
[45] &  &  &  &  &  &  &  & C\end{array}\right),
\label{u1}
\eea
where
we introduce some quantities that are functions of $\la, \la', \rho,
\alpha, \phi$, and $\varphi$ with (\ref{pqrt1})
as follows:
\bea
&& A \equiv p^{3},\qquad B \equiv pq^{2},\qquad C \equiv pr^{2},\qquad D \equiv
pt^{2}, \qquad E \equiv \la' qrt,\qquad F \equiv \rho qrt,
\nonu \\
&& D_{1} \equiv
\frac{D(K_{2}^{2}\rho^{2}+K_{3}^{2}\la'^{2})}{K_{2}^{2}K_{3}^{2}G^{2}},
\qquad 
D_{2} \equiv \frac{D(K_{3}^{2}\rho^{2}+K_{2}^{2}\la'^{2})}{G^{2}},\qquad
D_{3} \equiv \frac{(K_{3}^{2}-K_{2}^{2})D\rho\la'}{K_{2}K_{3}G^{2}},\nonu \\
&& G \equiv \sqrt{\la'^{2}+\rho^{2}}, \qquad
K_{1} \equiv e^{i\alpha},\qquad K_{2} \equiv e^{i\phi},\qquad 
K_{3} \equiv e^{i\varphi},
\label{Change}
\eea
and
\bea
v_{1}  =  \left(\begin{array}{ccccc}
 & [12] & [34] & [56] & [78]\\{}
[12] & \frac{L}{K_{1}} & \frac{M}{K_{1}} & \frac{M}{K_{1}} & K_{1}M\\{}
[34] &  & \frac{L}{K_{1}} & \frac{M}{K_{1}} & K_{1}M\\{}
[56] &  &  & \frac{L}{K_{1}} & K_{1}M\\{}
[78] &  &  &  & K_{1}^{3}L\end{array}\right),
\nonu 
\eea
\bea
v_{2} =  \left(\begin{array}{ccccccccc}
 & [13] & [24] & [57] & [68] & [14] & [23] & [58] & [67]\\{}
[13] & \frac{P}{K_{1}} & -\frac{N}{K_{1}} & -\frac{R}{K_{2}G} &
 \frac{K_{2}R}{G} 
& 0 & 0 & -\frac{K_{3}S}{G} & -\frac{S}{K_{3}G}\\{}
[24] &  & \frac{P}{K_{1}} & \frac{R}{K_{2}G} & -\frac{K_{2}R}{G} & 0 &
 0 & 
\frac{K_{3}S}{G} & \frac{S}{K_{3}G}\\{}
[57] &  &  & K_{1}P_{1} & -K_{1}N & -\frac{S}{K_{3}G} &
 -\frac{S}{K_{3}G} & 
K_{1}P_{3} & 0\\{}
[68] &  &  &  & K_{1}P_{2} & \frac{K_{3}S}{G} & \frac{K_{3}S}{G} & 0 & K_{1}P_{3}\\{}
[14] &  &  &  &  & \frac{P}{K_{1}} & \frac{N}{K_{1}} & \frac{K_{2}R}{G} & \frac{R}{K_{2}G}\\{}
[23] &  &  &  &  &  & \frac{P}{K_{1}} & \frac{K_{2}R}{G} & \frac{R}{K_{2}G}\\{}
[58] &  &  &  &  &  &  & K_{1}P_{2} & K_{1}N\\{}
[67] &  &  &  &  &  &  &  & K_{1}P_{1}\end{array}\right),
\nonu 
\eea
\bea
v_{3}  =  \left(\begin{array}{ccccccccc}
 & [15] & [26] & [37] & [48] & [16] & [25] & [38] & [47]\\{}
[15] & \frac{P}{K_{1}} & -\frac{N}{K_{1}} & \frac{R}{K_{2}G} &
 -\frac{K_{2}R}{G} & 0 & 0 
& \frac{K_{3}S}{G} & \frac{S}{K_{3}G}\\{}
[26] &  & \frac{P}{K_{1}} & -\frac{R}{K_{2}G} & \frac{K_{2}R}{G} & 0 &
 0 & 
-\frac{K_{3}S}{G} & -\frac{S}{K_{3}G}\\{}
[37] &  &  & K_{1}P_{1} & -K_{1}N & \frac{S}{K_{3}G} & \frac{S}{K_{3}G} & K_{1}P_{3} & 0\\{}
[48] &  &  &  & K_{1}P_{2} & -\frac{K_{3}S}{G} & -\frac{K_{3}S}{G} & 0 & K_{1}P_{3}\\{}
[16] &  &  &  &  & \frac{P}{K_{1}} & \frac{N}{K_{1}} & -\frac{K_{2}R}{G} & -\frac{R}{K_{2}G}\\{}
[25] &  &  &  &  &  & \frac{P}{K_{1}} & -\frac{K_{2}R}{G} & -\frac{R}{K_{2}G}\\{}
[38] &  &  &  &  &  &  & K_{1}P_{2} & K_{1}N\\{}
[47] &  &  &  &  &  &  &  & K_{1}P_{1}\end{array}\right),
\nonu 
\eea
\bea
v_{4}  =  \left(\begin{array}{ccccccccc}
 & [17] & [28] & [35] & [46] & [18] & [27] & [36] & [45]\\{}
[17] & K_{1}P_{1} & -K_{1}N & -\frac{R}{K_{2}G} & \frac{R}{K_{2}G} &
 K_{1}P_{3} & 0 
& -\frac{S}{K_{3}G} & -\frac{S}{K_{3}G}\\{}
[28] &  & K_{1}P_{2} & \frac{K_{2}R}{G} & -\frac{K_{2}R}{G} & 0 &
 K_{1}P_{3} & 
\frac{K_{3}S}{G} & \frac{K_{3}S}{G}\\{}
[35] &  &  & \frac{P}{K_{1}} & -\frac{N}{K_{1}} & -\frac{K_{3}S}{G} & 
-\frac{S}{K_{3}G} & 0 & 0\\{}
[46] &  &  &  & \frac{P}{K_{1}} & \frac{K_{3}S}{G} & \frac{S}{K_{3}G} & 0 & 0\\{}
[18] &  &  &  &  & K_{1}P_{2} & K_{1}N & \frac{K_{2}R}{G} & \frac{K_{2}R}{G}\\{}
[27] &  &  &  &  &  & K_{1}P_{1} & \frac{R}{K_{2}G} & \frac{R}{K_{2}G}\\{}
[36] &  &  &  &  &  &  & \frac{P}{K_{1}} & \frac{N}{K_{1}}\\{}
[45] &  &  &  &  &  &  &  & \frac{P}{K_{1}}\end{array}\right),
\label{v1}
\eea
where
we introduce some quantities that are functions of $\la, \la', \rho,
\alpha, \phi$, and $\varphi$ with (\ref{pqrt1})
as follows:
\bea
&& L \equiv q^{3},\qquad M \equiv p^{2}q,\qquad N \equiv qr^{2},\qquad P \equiv
qt^{2}, \qquad
R \equiv \la' prt,\qquad S \equiv \rho prt, \nonu \\
&& P_{1} \equiv 
\frac{P(K_{2}^{2}\rho^{2}+K_{3}^{2}\la'^{2})}{K_{2}^{2}K_{3}^{2}G^{2}},\qquad 
P_{2} \equiv \frac{P(K_{2}^{2}\la'^{2}+K_{3}^{2}\rho^{2})}{G^{2}},\quad 
P_{3}\equiv 
\frac{(K_{3}^{2}-K_{2}^{2})P\rho\la'}{K_{2}K_{3}G^{2}}.
\label{Change1}
\eea
The lower triangle part of (\ref{u1}) and (\ref{v1}) 
can be read off from the upper triangle part by
hermitian property.
Also, the other kinds of 
28-beins $u_{IJ}^{\;\;\;KL}$ and $v_{IJKL}$ are obtained by taking a
complex conjugation of (\ref{u1}) and (\ref{v1}). 
The complex conjugation operation 
can be done by raising or lowering the indices:  
$(u_{IJ}^{\;\;\;KL})^{\ast} =  u^{IJ}_{\;\;\;KL}$ and so on using 
(\ref{uv1}), (\ref{u1}), (\ref{v1}), (\ref{Change}) and (\ref{Change1}).
Note that under the relations (\ref{concon}) together with corresponding 
relations of $\theta$ and $\psi$ to the variables of section 3, 
the above (\ref{u1}) and
(\ref{v1})
are related to (\ref{u}) and (\ref{v}) each other.

The final expression of kinetic terms
is given by
\bea
\frac{1}{96}|A_{\mu}^{\; ijkl}|^{2} & = &
\frac{1}{16(\rho^{2}+\la'^{2})^{2}}\left[-8 \rho^{4}(\pa_{\mu}\varphi)^{2}-
5\rho^{2}\la'^{2}(\pa_{\mu}\phi)^{2}+6\rho^{2}
\la'^{2}(\pa_{\mu}\phi)(\pa_{\mu}\varphi)
-5\rho^{2}\la'^{2}(\pa_{\mu}\varphi)^{2}\right.\nonu \\
 &   -& 
8\la'^{4}(\pa_{\mu}\phi)^{2}+48(\rho^{2}+\la'^{2})^{2}(\pa_{\mu}\la)^{2}
-6(\rho^{2}+\la'^{2})^{2}(\pa_{\mu}\alpha)^{2}
\nonu \\
 &   +&2\rho\la'
(5+64\rho^{2}+64\la'^{2})(\pa_{\mu}\la')(\pa_{\mu}\rho)+(64
\rho^{4}+(-5+64\rho^{2})\la'^{2}) (\pa_{\mu}\rho)^{2}\nonu \\
 &  
 +&(64\la'^{4}+\rho^{2}(-5+64\la'^{2}))(\pa_{\mu}\la')^{2}+
6(\rho^{2}+\la'^{2})^{2}\cosh4\lambda (\pa_{\mu}\alpha)^{2} \nonu\\
 &  
 +& 4(\rho^{2}(\pa_{\mu}\la')^{2}+2\rho^{4}(\pa_{\mu}\varphi)^{2}-
2\rho\la'(\pa_{\mu}\la')(\pa_{\mu}\rho)+\la'^2((\pa_{\mu}\rho)^{2}+
\rho^2 (\pa_{\mu}\phi-\pa_{\mu}\varphi)^{2}) \nonu\\
 & + &  2\la'^4(\pa_{\mu}\phi)^{2})\cosh(4\sqrt{\rho^{2}+
\la'^{2}})+(\rho^2(\pa_{\mu}\la')^{2}-2\rho\la'(\pa_{\mu}\la')
(\pa_{\mu}\rho) \nonu\\
 & + & \la'^2
((\pa_{\mu}\rho)^{2}+\rho^2(\pa_{\mu}\phi+\pa_{\mu}\varphi)^{2}
))
\cosh(8\sqrt{\rho^{2}+\la'^{2}})\nonu\\
 &  
 +& 8((\rho^2(\pa_{\mu}\la')^{2}-2\rho\la'(\pa_{\mu}\la')(\pa_{\mu}\rho)+
\la'^2((\pa_{\mu}\rho)^{2}-\rho^2(\pa_{\mu}\phi+\pa_{\mu}\varphi)^{2}
))\cos2(\phi-\varphi) \nonu\\
 & - &
 \left.2\rho\la'(\pa_{\mu}\phi+\pa_{\mu}\varphi)(\rho\pa_{\mu}\la'-
\la'\pa_{\mu}\rho)\sin2(\phi-\varphi))
\sinh^4(2\sqrt{\rho^{2}+\la'^{2}})\right].
\label{kinetic2}
\eea
In particular, this becomes very simple when $\varphi=\phi$ and it
leads to
\bea
&&\frac{3}{8}\left(
\partial_{\mu} \lambda \right)^2 +\frac{3}{4} \sinh^2
\left(\frac{\la}{\sqrt{2}}\right) 
\left(
\partial_{\mu} \alpha \right)^2  +
\frac{1}{2}  \left( \partial_{\mu} \sqrt{\la'^2+\rho^2 }\right)^2 +
 \sinh^2 \left(\frac{\sqrt{\la'^2+\rho^2}}{\sqrt{2}} \right) 
\left( \partial_{\mu} \phi \right)^2
 \nonu \\
&&+ 
\frac{\sinh^2(\sqrt{2} \sqrt{\la'^2+\rho^2})}{4(\la'^2+\rho^2)}\left(
  \frac{\la'\pa_{\mu} \rho- \rho \pa_{\mu} \la'} {\sqrt{\la'^2+\rho^2}}\right)^2.
\label{simplekin}
\eea


\end{document}